\documentclass[12pt]{article}
\usepackage{bbm}
\usepackage{epsfig}
\usepackage{array}
\usepackage{float}
\usepackage{axodraw}

\usepackage{a4}

\usepackage{a4wide}
\usepackage{wasysym}

\def\gs{\mathrel{
   \rlap{\raise 0.511ex \hbox{$>$}}{\lower 0.511ex \hbox{$\sim$}}}}
\def\ls{\mathrel{
   \rlap{\raise 0.511ex \hbox{$<$}}{\lower 0.511ex \hbox{$\sim$}}}}
\newcommand{\obb}{0\mbox{$\nu\beta\beta$}}
\newcommand{\onbb}{neutrinoless double beta decay}
\newcommand{\ba}{\begin{array}{c}}
\newcommand{\baz}{\begin{array}{cc}}
\newcommand{\bad}{\begin{array}{ccc}}
\newcommand{\bav}{\begin{array}{cccc}}
\newcommand{\baf}{\begin{array}{ccccc}}
\newcommand{\bea}{\begin{equation} \begin{array}{c}}
\newcommand{\eea}{ \end{array} \end{equation}}
\newcommand{\ea}{\end{array}}

\newcommand{\dms}{\mbox{$\Delta m^2_{\odot}$}}
\newcommand{\dma}{\mbox{$\Delta m^2_{\rm A}$}}
\newcommand{\lsndo}{\mbox{$\Delta m^2_{\rm s1}$}}
\newcommand{\lsndt}{\mbox{$\Delta m^2_{\rm s2}$}}
\newcommand{\lsndth}{\mbox{$\Delta m^2_{\rm s3}$}}

\newcommand{\lsndot}{\mbox{$\Delta \tilde m^2_{\rm s1}$}}
\newcommand{\lsndtt}{\mbox{$\Delta \tilde m^2_{\rm s2}$}}
\newcommand{\meff}{\mbox{$\langle m \rangle$}}

\newcommand{\be}{\begin{eqnarray}}
\newcommand{\ee}{\end{eqnarray}}

\newcommand{\sss}{\sin^2 \theta_{\odot}}
\newcommand{\sch}{\sin^2 \theta_{\rm CHOOZ}}

\newcommand{\css}{\cos^2 \theta_{\odot}}

\begin{document}

\title{\vspace{-1cm}
\vspace{-0.3cm}
\hfill {\small arXiv:0706.1462} 
\vskip 0.5cm
\bf \large 
MiniBooNE Results and Neutrino Schemes with 2 sterile 
Neutrinos: Possible Mass Orderings and Observables related to 
Neutrino Masses
}
\author{
Srubabati Goswami$^{a,b}$\,\thanks{email: \tt sruba@mri.ernet.in}~~~and~~
Werner Rodejohann$^b$\,\thanks{email: \tt werner.rodejohann@mpi-hd.mpg.de} 
\\\\
$^a${\normalsize \it Harish--Chandra Research Institute, Chhatnag Road,}\\
{\normalsize \it Jhunsi, Allahabad 211 019, India}\\ \\
$^b${\normalsize \it Max--Planck--Institut f\"ur Kernphysik,}\\
{\normalsize \it Postfach 10 39 80, D-69029 Heidelberg, Germany}
}
\date{}
\maketitle
\thispagestyle{empty}
\begin{abstract}

\noindent 
The MiniBooNE and LSND experiments are compatible with each other 
when two sterile neutrinos are added to the three active ones. 
In this case there are eight possible mass orderings. 
In two of them both sterile neutrinos are 
heavier than the three active ones. 
In the next two scenarios both sterile neutrinos are 
lighter than the three active ones.  
The remaining four scenarios have 
one sterile neutrino heavier 
and another lighter than the three active ones. 
We analyze all scenarios with respect to their 
predictions for mass-related observables. These are 
the sum of neutrino masses as constrained by cosmological 
observations, the kinematic mass parameter as measurable in the 
KATRIN experiment, and the effective mass governing  
neutrinoless double beta decay. 
It is investigated how these non-oscillation probes 
can distinguish between the eight scenarios. Six of the eight possible 
mass orderings predict positive signals in the KATRIN and future neutrinoless 
double beta decay experiments. 
We also remark on scenarios with three sterile neutrinos. 
In addition we make some comments on the possibility of using 
decays of high energy astrophysical neutrinos to discriminate 
between the mass orderings in presence of two sterile 
neutrinos.

\end{abstract}
\newpage

\section{\label{sec:intro}Introduction}
The long awaited results of the MiniBooNE experiment \cite{finally} 
showed that scenarios in which one sterile neutrino is added to the 
three active ones are incompatible with data. 
Such schemes were motivated by the results from the LSND experiment 
\cite{lsnd}, which observed flavor transitions interpreted as 
$\bar{\nu}_\mu \leftrightarrow \bar{\nu}_e$ neutrino oscillations. 
A number of authors 
\cite{4mix,4cosmo,40vbb,BPP,GR} investigated the 
implications of such schemes. 
In particular, it was realized \cite{vallesterile} that so-called 2+2 
scenarios (two pairs of neutrinos 
close in mass separated by a large gap) are ruled out and that only 
3+1 scenarios (three mostly active neutrinos separated by a large gap 
from the mostly sterile one) are allowed, though only small part 
of the parameter space survived. The MiniBooNE results 
ruled out even this part \cite{3+2data} 
by excluding the LSND parameter space at the 98 $\%$ C.L.~\cite{finally}. 

Allowing one more sterile neutrino to enter the stage 
improves the compatibility of LSND with other experiments \cite{3+20,3+2CP} 
and in particular renders the MiniBooNE and LSND 
experiments compatible \cite{3+2data}. 
Only comparably few models for neutrino schemes with two extra 
sterile neutrinos have been constructed \cite{3+2mod}, 
and the potentially rich 
phenomenology of such scenarios is hardly investigated 
\cite{3+2phen0,3+2phen1}.\\

With two sterile neutrinos added to the usual three, one has 
eight possible mass orderings, which should be compared with the 
two schemes (normal and inverted ordering) in case of 
``only'' three active neutrinos. 
We study in this paper the predictions of the eight cases for 
mass-related observables. 
We investigate the sum of neutrino masses 
as constrained by  cosmological 
observations, the kinematic mass parameter as measurable in the 
KATRIN experiment (partly analyzed also 
in Ref.~\cite{3+20}), and the effective mass controlling  
neutrinoless double beta decay\footnote{For related 
analyzes in different 
sterile neutrino scenarios, see, e.g., Ref.~\cite{others}.}. 
We also investigate how and if mass-related observables can contribute 
to distinguish the possibilities. 
The mass patterns in the 3+2 scheme can be classified 
in three main classes: 
\begin{itemize}
\item two 2+3 scenarios: the 
two sterile neutrinos are heavier than the three 
active ones; 
\item two 3+2 scenarios: the 
two sterile neutrinos are lighter than the three 
active ones; 
\item four 1+3+1 scenarios: one sterile neutrino is heavier than the three 
active ones which in turn are heavier than the second sterile neutrino. 
\end{itemize}

The paper is build up as follows: first we summarize the required 
formalism in Section \ref{sec:form} before outlining 
in Section \ref{sec:schemes} the 
eight possible mass orderings for scenarios with two sterile 
neutrinos. In Section 
\ref{sec:3l2h} we study the mass-related observables for the two 
scenarios which have the sterile 
neutrinos heavier than the active ones, while 
in Section \ref{sec:3h2l} the two scenarios in which the sterile 
neutrinos are lighter than the active neutrinos are analyzed. 
Section \ref{sec:131} contains the four 1+3+1 
scenarios and a short discussion on scenarios with three 
sterile neutrinos is delegated to Appendix \ref{sec:3+3}. 
In Appendix \ref{sec:UHE} we discuss another 
interesting possibility to 
distinguish between the different scenarios outlined above at 
neutrino telescopes, allowing high energy astrophysical 
neutrinos to decay.
Finally, in Section \ref{sec:concl} we discuss and summarize our 
findings.

\section{\label{sec:form}Formalism}

\subsection{\label{sec:mix}Neutrino Mixing}
Neutrino mixing is described by the leptonic mixing, 
or Pontecorvo-Maki-Nakagawa-Sakata (PMNS) matrix $U$: 
 \bea \label{eq:Upara}
U = 
\left( 
\baf  
U_{e1} & U_{e2} & U_{e3} & U_{e4} & U_{e5} \\
U_{\mu 1} & U_{\mu 2} & U_{ \mu 3} & U_{ \mu 4} & U_{ \mu 5}\\
U_{\tau 1} & U_{\tau 2} & U_{\tau 3} & U_{\tau 4} & U_{ \tau 5} \\ 
U_{s_1 1} & U_{s_1 2} & U_{s_1 3} & U_{s_1 4} & U_{s_1 5} \\
U_{s_2 1} & U_{s_2 2} & U_{s_2 3} & U_{s_2 4} & U_{s_2 5}
\ea 
\right)
~. 
\eea
It links the mass eigenstates $\nu_{1,2,3,4,5}$ having 
masses $m_{1,2,3,4,5}$ with the three active 
flavor states $\nu_{e, \mu, \tau}$ and the two new sterile states 
$\nu_{s_1}$  and $\nu_{s_2}$. 
In what follows we denote the mostly active neutrinos with 
$\nu_{1,2,3}$. Regardless of their ordering (normal or inverted) 
we have 
\be \label{eq:3mix}
|U_{e1}|^2 \simeq \cos^2 \theta_{\odot}\,,~ 
|U_{e2}|^2 \simeq \sin^2 \theta_{\odot} ~\mbox{ and }~
|U_{e3}|^2 \simeq \sin^2 \theta_{\rm CHOOZ}~, 
\ee
where $\theta_{\odot}$ is the mixing angle for solar and KamLAND 
neutrinos and $\theta_{\rm CHOOZ}$ the mixing angle for short baseline 
reactor neutrinos. 
We further have the mass-squared differences governing 
solar and atmospheric neutrino oscillations. In the following, we will 
use the following best-fit and $3\sigma$ ranges \cite{GonMal}: 
\bea \label{eq:data}
\sin^2 \theta_{\odot} = 0.30^{+0.08}_{-0.05} ~~\mbox{ with }~
\dms = \left(8.0^{+1.0}_{-1.0} \right) \cdot 10^{-5}~{\rm eV}^2~,\\[0.2cm]
\sin^2 \theta_{\rm CHOOZ} = 0.00^{+0.04}_{-0.0}~~\mbox{ with }~
\dma = \left(2.6^{+0.6}_{-0.6}\right) \cdot 10^{-3}~{\rm eV}^2~. 
\eea
The related typical 
mass scales are therefore 
$\sqrt{\dms} \simeq 0.009$ eV and $\sqrt{\dma} \simeq 0.05$ eV, 
respectively. 
In what regards the two additional sterile neutrinos, 
the analysis in Ref.~\cite{3+2data} resulted in the following 
best-fit values\footnote{We consider only the analysis 
which uses the MiniBooNE results above reconstructed neutrino 
energies of 475 MeV, because results from the lowest energy bin are 
not well understood \cite{finally}.} 
\bea \label{eq:stBF}
\lsndo = 6.49^{+1.0}_{-1.0}~{\rm eV}^2~~\mbox{ with }~ 
|U_{e 5}| = 0.12~,\\[0.2cm]
\lsndt = 0.89^{+0.1}_{-0.1} ~{\rm eV}^2~~ 
\mbox{ with }~ |U_{e 4}| = 0.11~.
\eea
In what follows we will give for our observables explicit numerical 
values obtained with not only these best-fit values, but 
also for another typical illustrative point in the parameter space: 
\bea \label{eq:st2}
\lsndo = 1.90^{+0.60}_{-0.90}~{\rm eV}^2~~\mbox{ with } ~
|U_{e 5}| = 0.12
~,\\[0.2cm]
\lsndt = 0.90
^{+0.05}_{-0.20}
~{\rm eV}^2~~ \mbox{ with } ~
|U_{e 4}| = 0.11~.
\eea
The main feature of this point is of course the smaller overall 
neutrino mass it implies. 
It corresponds approximately to the center of another, isolated region 
allowed at 90 $\%$ C.L.~of Figure 6 from Ref.~\cite{3+2data}. 
The two central points from Eqs.~(\ref{eq:stBF}) and (\ref{eq:st2}) 
are quite typical for the situation in the presence of two 
sterile neutrinos and we will make frequent use of them. 
With the values of the two new mass-squared differences we can 
estimate typical neutrino mass scales, 
which we will encounter frequently in the 
following: $\sqrt{\lsndo} \simeq 2.55~(1.38)$ eV, 
$\sqrt{\lsndt} \simeq 0.94~(0.95)$ eV, 
$\sqrt{\lsndo - \lsndt} \simeq 2.37~(1.00)$ eV, and 
$\sqrt{\lsndo + \lsndt} \simeq 2.72~(1.67)$ eV. 

In Ref.~\cite{3+2data} the allowed ranges of the mixing matrix elements 
$|U_{e4}|$ and $|U_{e5}|$ are not given. However, we will see in our 
discussions that especially for neutrinoless double beta decay 
it is sometimes important to analyze the impact 
of varying these parameters. 
In absence of any information we have varied $|U_{e4}|$ and $|U_{e5}|$ 
by 50\% around the best-fit points in Eqs.~(\ref{eq:stBF}) and 
(\ref{eq:st2}). Thus we consider the following ranges 
for the parameters $|U_{e4}|$ and $|U_{e5}|$:
\be
|U_{e 5}| = 0.12
^{+0.06}_{-0.06} ~~\mbox{ and } ~~ |U_{e 4}| = 0.11^{+0.05}_{-0.05} ~.
\label{eq:rangeUe45}
\ee
for both the best-fit point and the second illustrative point 
considered in Eq.~(\ref{eq:st2}).

\subsection{\label{sec:mass}Neutrino Masses}
As neutrino oscillations are sensitive only to mass-squared 
differences, the neutrino mass scale is not known, but only 
limited from above by different experiments and observations. 
Typically, the mass scale is inversely 
proportional to the scale of the mechanism which is responsible 
for neutrino mass. Therefore, knowing the mass is a 
very important step towards the understanding of neutrino physics.

Mass-related observables are the sum of neutrino masses 
\be \label{eq:Sigma}
\sum = \sum_i m_i~,
\ee
which can be inferred from cosmological observations. 
Typical limits are smaller than about 1 eV \cite{steen,raffelt06,lyman,nucosmo}, but they 
depend on the used data sets, the number of neutrino species and 
how the mass is distributed among the different neutrinos. We will 
discuss this in more detail in Section \ref{sec:concl}.    
One also has the kinematic neutrino mass parameter 
measurable in 
$\beta$-decay experiments  
\be \label{eq:KATRIN}
m_\beta = \sqrt{\sum_i |U_{ei}|^2 \, m_i^2 }~. 
\ee
This quantity is measured when the electron 
energy interval around the endpoint of the 
investigated beta decay is much larger than the $m_i$, otherwise 
corrections to this formula are required \cite{curiespec}.
 \begin{table}[t]\hspace{-.6cm}
\begin{tabular}{|c|c|c|c|}\hline 
\small scheme & \small $\Sigma$ & \small $m_\beta$ 
& \small \meff \\ \hline \hline
\small NH & \small $\sqrt{\dma} $ 
& \small $\sqrt{\sss \, \dms + \sch \, \dma}  $
& \small $\left| \sss \, \sqrt{\dms} + \sch \, \sqrt{\dma} \, 
e^{i \alpha_{32}} \right|$ \\ \hline 
\small IH & \small $2 \, \sqrt{\dma}$ 
& \small $\sqrt{\dma}$ & \small $\sqrt{\dma} \, 
\sqrt{1 - \sin^2 2 \theta_\odot \, \sin^2 \alpha_2/2}$ \\ \hline 
\small QD & \small $3 \, m_0$ & \small $m_0$ 
& \small $m_0 \, 
\sqrt{1 - \sin^2 2 \theta_\odot \, \sin^2 \alpha_2/2}$ \\ \hline 
\end{tabular}
\caption{\label{tab:3mass}Extreme limits of 3-flavor scenarios 
and the resulting mass-related observables. We have defined 
$\alpha_{32} = \alpha_3 - \alpha_2$.}
\end{table}
This condition is fulfilled for the values of the neutrino masses 
we choose\footnote{In principle, the analysis of Ref.~\cite{3+2data} 
allows sterile neutrino mass values of  $\Delta m^2 \ge 10$ 
eV$^2$, which are indeed 
close to the energy interval used by the upcoming KATRIN experiment 
\cite{KATRIN}. 
These values are however in very strong conflict with all 
mass-related observables  
and we therefore omit them. For the values 
used we estimate corrections to Eq.~(\ref{eq:KATRIN}) 
to be at most of order 10\%.}. The current limit on $ m_\beta$ 
is 2.2 eV at 95\% C.L. \cite{mainz}, 
and improvement by one order of magnitude is expected by the KATRIN 
experiment \cite{KATRIN}. 
Finally, we have the effective mass 
in neutrinoless double beta decay: 
\be \label{eq:meff}
\meff = \left| \sum_i U_{ei}^2 \, m_i \right| = 
\left|\sum_i |U_{ei}|^2 \, e^{i \alpha_i} \, m_i  \right| ~. 
\ee
Here $\alpha_{2,3,4,5}$ are the four possible and 
unknown\footnote{In fact, the analysis of 
neutrino scenarios with two sterile neutrinos gives some 
constraint on a $CP$ phase \cite{3+2data} 
(see also \cite{3+2CP}). 
Being a ``Dirac-phase'', however, it does not appear in \meff.}
Majorana phases (we can choose $\alpha_1 = 0$). 
Applying nuclear matrix element uncertainties 
on the current $90 \%$ C.L.~limits on lifetimes  \cite{0vbb1,0vbb2} 
gives limits on the effective mass in the range of 1 eV, 
and sizable improvement is expected also in this field \cite{APS_0vbb}. 

Let us summarize for the sake of comparison 
the situation in 3-flavor scenarios 
(see for instance \cite{mass_rec,meff_spec}). 
We have three extreme cases of the mass ordering, 
the normal hierarchy 
(NH, $m_3^2 \simeq \dma \gg m_2^2 \simeq \dms \gg m_1^2$), 
the inverted hierarchy 
(IH, $m_2^2 \simeq m_1^2 \simeq \dma \gg m_3^2$) 
and quasi-degenerate neutrinos 
(QD, $m_3^2 \simeq m_2^2 \simeq m_1^2 \equiv m_0^2 \gg \dma, \dms$). 
Table \ref{tab:3mass} shows the results for the 
mass-related observables. 
Obviously, $m_\beta$ is unobservably low  
for NH and IH, while $\meff$ and $\Sigma$ are for NH. 
It may be possible to probe the inverted hierarchy 
regime through future cosmological observations \cite{steen}. \\ 

Certain mass orderings to be discussed in the following will 
have problems with some of the three observables 
$\Sigma$, $m_\beta$ or $\meff$. Does this mean that 
they are ruled out? Not necessarily, because cosmological neutrino 
mass \cite{cosmo2} (and number \cite{cosmo}) 
limits can in principle be evaded, 
or relaxed by a factor of a few, by means of 
unknown neutrino interactions or other cosmological features. 
Furthermore, confronting a mass ordering with 
the limit on the effective mass makes only sense when neutrinos 
are Majorana particles, which however is a very well justified 
assumption. Only the kinematic parameter $m_\beta$ 
does not suffer from any underlying model assumption and provides an 
unambiguous test. We will leave aside discussions of the validity 
of the different limits in particular on $\Sigma$. 
Our aim is simply to study the 
predictions for the observables for all eight possible mass orderings, 
which we will outline in the next Section. 

\section{\label{sec:schemes}Eight Possible Mass Orderings in Neutrino 
Scenarios with two sterile Neutrinos}
As mentioned in the Introduction, analyzes of the LSND and MiniBooNE, 
as well as various other experiments, give a consistent picture only 
if two additional independent mass-squared differences 
$\lsndo> \lsndt$ are present. 
We assume here that 
the difference between the two, 
$\Delta m^2_{ss} \equiv \lsndo - \lsndt$, is much larger than \dma, 
an assumption used in the 3+2 analyzes in \cite{3+2data} from where 
we take the values of the additional parameters.

Our convention is the following: we call the masses of 
the three predominantly active neutrinos $m_3$, $m_2$ and $m_1$. The 
mixing among them is responsible for the solar, atmospheric 
and short baseline reactor neutrino oscillation results.  
If they are normally ordered, then $m_3> m_2> m_1$ with 
\be \label{eq:3NO}
m_3^2 - m_2^2 = \dma ~~\mbox{ and } ~~m_2^2 - m_1^2 = \dms~.
\ee 
They can also be inversely ordered, in which case $m_2> m_1> m_3$ and  
\be \label{eq:3IO}
m_1^2 - m_3^2 = \dma ~~\mbox{ and } ~~m_2^2 - m_1^2 = \dms~.
\ee 
We have to add two predominately sterile neutrino states, 
whose masses we denote by 
$m_4$ and $m_5$. 
They can either be heavier or lighter than the three 
active neutrinos (in what follows, we will omit for simplicity 
the word ``predominately'' or ``mostly'' in 
front of ``sterile neutrinos'' and ``active neutrinos''). 
Without loss of generality, we can choose that $m_5$ is either 
the largest or the smallest mass and associate it always with 
\lsndo, while $m_4$ is always associated with \lsndt. 
In this case, we do not have to rename the 
matrix elements $U_{e4}$ and $U_{e5}$, which quantify the 
mixing of the two additional neutrinos with the electron neutrino. 
In general, one could choose the labels of the masses such that 
always $m_5> m_4> m_3> m_2> m_1$ holds. In this case, however, 
the mixing matrix elements would be different for each of the 
possible mass orderings. With our convention, the values of $|U_{ei}|$ 
are fixed by Eqs.~(\ref{eq:3mix}) and (\ref{eq:stBF}, \ref{eq:st2}) 
and do not have to be relabeled.

Let us first discuss the case of {\it both} sterile neutrinos being either 
heavier or lighter than the three active ones:  
the largest of the two independent new mass-squared 
differences, \lsndo, is then always the largest possible mass-squared 
difference. If the two sterile states are above the three active ones 
(``2+3 scenarios''), then \lsndt~is the 
mass-squared difference between the lightest 
sterile state (which is also the second heaviest state) 
and the lightest available state (which is active). 
If the two sterile states are 
below the three active ones (``3+2 scenarios''), 
then \lsndt~is the mass-squared 
difference between the heaviest state (which is active) and the 
heaviest sterile state (which is the second lightest state). 
The four possible schemes are shown in Figs.~\ref{fig:3l2h} and 
\ref{fig:3h2l}. 
The names of the schemes are defined as follows: 
depending on whether the active neutrinos are 
normally or inversely ordered, the scheme has the capital letter ``N'' 
or ``I'' in its name. Depending on whether the two sterile 
neutrino masses $m_4$ and $m_5$ are lighter or heavier than the 
active ones, the capital letters ``SS'' appear after or in front of  
this capital letter. For instance, 
if $m_5> m_4> m_3> m_2> m_1$, then we call the scenario SSN, while 
for $m_2> m_1> m_3> m_4> m_5 $ we call it ISS. 

The last possible class of mass orderings is when one sterile neutrino 
is heavier then the three active ones, which in turn are heavier than the 
second sterile neutrino (``1+3+1 scenarios''). The heaviest 
neutrino can either be separated by \lsndo~or \lsndt~from the active 
neutrinos. If it is separated by \lsndo~(\lsndt) then we call the scenario 
SNSa or SISa (SNSb or SISb). 
The possibilities are shown in Fig.~\ref{fig:131}. We note here that 
the fit of Ref.~\cite{3+2data}, and also the analyzes of 
Refs.~\cite{3+20,3+2CP}, do strictly speaking not apply to these 
schemes. The reason is that in the oscillation probabilities for 
$\nu_\mu \rightarrow \nu_e$ transitions there 
are not only terms proportional to $\sin^2 \lsndo \, \frac{L}{4 E}$ and 
to $\sin^2 \lsndt \, \frac{L}{4 E}$, but also an interference term 
proportional to $\cos (\Delta \tilde m^2 \, \frac{L}{4 E} + \delta)$, where 
$\delta$ is a CP phase (which does not appear in survival probabilities). 
Refs.~\cite{3+2data,3+20,3+2CP} make   
the implicit assumption $|\Delta \tilde m^2| = |\lsndo| - |\lsndt|$, 
which is not fulfilled for the 1+3+1 scenarios. In lack of any 
detailed fit of the data within these schemes we will assume for simplicity 
that the mass-squared differences are the same. The values of the 
mass-squared differences resulting from fits taking into account the 
1+3+1 case will remain of course in the eV range.\\
 
Summarizing we end up with eight different schemes\footnote{In 
scenarios with three sterile neutrinos one would have 
16 possible mass orderings, see Appendix \ref{sec:3+3}.}.  
We can already at the present stage make some general statements. 
First of all, the sum of neutrino masses depends basically 
only on the new mass-squared differences and typical values will be  
\be
\Sigma \ge \sqrt{\lsndo} + \sqrt{\lsndt}~~ \mbox{ or } ~
\Sigma \ge 3 \, \sqrt{\lsndo} ~~\mbox{ or } ~
\Sigma \ge 3 \, \sqrt{\lsndt}
~, 
\ee
depending on the details of the mass ordering. 
In general, all of them are expected to face serious 
problems with cosmology, 
and the smaller the mass-squared 
differences $\lsndo$ and $\lsndt$ are, the 
smaller $\Sigma$. 
Another point worth mentioning is that the effective 
mass governing \onbb~can be written as 
\bea
\meff = \left| \meff^{3} + \meff^{\rm st} \right|~,\\[0.2cm]
\mbox{ where  }~ \meff^{3} \equiv 
\cos^2 \theta_\odot \, m_1 + \sin^2 \theta_\odot \, m_2 \, e^{i \alpha_2} 
+ \sin^2 \theta_{\rm CHOOZ} \, m_3 \, e^{i \alpha_3} \\[0.2cm]
~~\mbox{ and }~ 
\meff^{\rm st} \equiv |U_{e4}|^2 \, m_4 \, e^{i \alpha_4}  
+ |U_{e5}|^2 \, m_5 \, e^{i \alpha_5} ~. 
\eea
Obviously, $|\meff^{3}|$ is an effective mass similar to the one 
analyzed in the usual three-flavor 
situation \cite{mass_rec}, cf.~Table \ref{tab:3mass}.  
The quantity $|\meff^{\rm st}|$ is the contribution from the two sterile 
states. We will encounter in the following all cases: 
dominance of the sterile contribution, dominance of the active 
contribution, and equal-sized contributions, leading potentially 
to complete cancellation. 
The same cases are also present for the kinematic neutrino 
mass $m_\beta$, where however no cancellation is possible as it 
is given by an incoherent sum.\\

We will discuss now the three mass-related observables for the 
eight possible mass orderings. 
We give approximate analytic expressions for these observables 
in the limit of vanishing smallest mass and $\sch$.  
We use these  expressions to give  illustrative 
numerical values  
in each case for the best-fit values of the oscillation 
parameters (or for the second illustrative central point 
from Eq.~(\ref{eq:st2})). 
We will also plot the observables as a function of the smallest neutrino 
mass for the central points as well as by varying the 
parameters in  their corresponding allowed ranges from 
Eqs.~(\ref{eq:data}, \ref{eq:stBF}, \ref{eq:st2}). 

\section{\label{sec:3l2h}Sterile Neutrinos heavier than 
active Neutrinos: 2+3 Scenarios}

\subsection{\label{sec:54N}Scheme SSN}
In this scheme, $m_5> m_4> m_3> m_2> m_1$. The three lowest states 
account for the solar and atmospheric neutrino mass-squared differences 
according to Eq.~(\ref{eq:3NO}). We have 
\be
\lsndo = m_5^2 - m_1^2 ~~\mbox{ and } ~~\lsndt = m_4^2 - m_1^2~.
\ee
Schematically, this scheme is shown in Fig.~\ref{fig:3l2h}.
We can express the individual masses in terms of the smallest mass 
$m_1$ and the independent mass-squared differences:
\bea \label{eq:54Nmass}
m_2 = \sqrt{\dms + m_1^2}\,,~m_3 = \sqrt{\dma + \dms + m_1^2}~,\\[0.2cm]
m_4 = \sqrt{\lsndt + m_1^2}\,,~m_5 = \sqrt{\lsndo + m_1^2}~.
\eea
The typical masses are therefore 
$m_2 \simeq \sqrt{\dms} \simeq 0.01$ eV, 
$m_3 \simeq \sqrt{\dma} \simeq 0.05$ eV, 
$m_4 \simeq \sqrt{\lsndt} \simeq 0.94~(0.95)$ eV and 
$m_5 \simeq \sqrt{\lsndo} \simeq 2.55~(1.38)$ eV. The three lightest 
neutrinos have the same values as in a normal hierarchical 3-flavor 
framework. The upper left plot in Fig.~\ref{fig:masses} shows the 
individual masses as a function of the smallest mass in this picture. 
The limit where all five neutrinos are quasi-degenerate  
comes when the smallest mass is beyond  1 eV. 

If this scheme is realized then the sum of neutrino masses 
which is constrained from  cosmology is given   
as   
\bea \label{eq:Sigma54N}
\Sigma^{\rm SSN} \simeq \sqrt{\dms} + \sqrt{\dma} + \sqrt{\lsndo} 
+ \sqrt{\lsndt} \\[0.2cm]
\simeq \sqrt{\lsndo} + \sqrt{\lsndt} \simeq 3.49~(2.32)~\rm eV ~.
\eea
Neglecting $m_1$ is a good approximation as long as $m_1 \ls 0.1$ eV. 
For such values, $\Sigma$ lies roughly between 3.3 and 3.8 (or 
1.9 and 2.6) eV, 
but it can reach unrealistically large values of 10 eV for a smallest 
mass in the eV range.
 
The kinematic mass is also mainly given by the sterile neutrino contribution:
\bea \label{eq:KATRIN54N}
m_\beta^{\rm SSN} \simeq \sqrt{\sss \, \dms + \sch \, \dma + 
|U_{e4}|^2 \, \lsndt + |U_{e5}|^2 \, \lsndo } \\[0.2cm]
\simeq \sqrt{|U_{e4}|^2 \, \lsndt + |U_{e5}|^2 \, \lsndo} 
\simeq 
0.32~(0.20)~\rm eV ~.
\eea 
Both $\Sigma$ and $m_\beta$ are shown in Fig.~\ref{fig:mass1}, 
where the solid lines are 
these quantities at the best-fit 
values of Eq.~(\ref{eq:stBF}) 
whereas the bands are obtained by varying the 
masses and the mixing angles within their allowed ranges. 
The KATRIN experiment, having a sensitivity of 0.3 eV, 
will find a positive signal if $m_1 \gs 0.3$ eV. For smaller 
values, however, $m_\beta$ can lie below 0.3 eV. If $m_\beta$ larger than 
0.5 eV is found then this scenario is ruled out, unless $m_1 \gs 0.2$ eV. 
No qualitatively different features are found for the second point from 
Eq.~(\ref{eq:st2}).

Finally, \onbb~should be triggered by an effective mass given by  
\bea \label{eq:meff54N}
\meff^{\rm SSN} \simeq 
\left| \sss \, \sqrt{\dms} + \sch \, \sqrt{\dma} \, 
e^{i (\alpha_3 - \alpha_2)} \right. \\[0.2cm]
\left. + |U_{e4}|^2 \, \sqrt{\lsndt} \, e^{i (\alpha_4 - \alpha_2)} 
+ |U_{e5}|^2 \, \sqrt{\lsndo} \, e^{i (\alpha_5 - \alpha_2)} \right| 
\\[0.2cm]
\simeq 
\left| |U_{e4}|^2 \, \sqrt{\lsndt} + 
|U_{e5}|^2 \, \sqrt{\lsndo} \, e^{i (\alpha_5 - \alpha_4)} \right| 
\simeq (0.025 \div 0.048)~\rm eV ~,
\eea 
where the two sterile neutrinos provide the leading contribution.
In case of the  second typical point from Eq.~(\ref{eq:st2}) the two 
additional neutrinos give a leading contribution between 
0.008 and 0.031 eV. The upper left panels of 
Figs.~\ref{fig:meff1} and \ref{fig:meff2} show the effective mass as 
a function of the smallest mass in this scenario.  
The shaded region inside is drawn for the best-fit values of mass and 
mixing parameters and varying the Majorana phases between 0 and $2 \pi$,  
while for the outer shaded regions we vary these parameters also 
in their permissible ranges. 
The figures show that 
neglecting  the smallest mass   
is a good approximation as long as $m_1 \ls 0.01$ eV. 
There is for both central points a cancellation regime 
(for $m_1$ between 0.02 and 0.1 eV) in which 
the effective mass vanishes or becomes unobservably small. 
If we use the ranges around the two central points of the sterile neutrino 
parameters, then the two terms can cancel when the 
conditions $|U_{e4}/U_{e5}|^2 = \sqrt{\lsndo/\lsndt}$ 
and $\alpha_5 - \alpha_4 \simeq \pi$ are fulfilled. 
For the smallest neutrino mass above 0.1 eV the effective mass 
cannot vanish due to the non-maximal solar neutrino mixing angle. 
For smaller values of $m_1$ the scenario is ruled out if 
\meff~is found to be larger than 0.1 eV.

\subsection{\label{sec:54I}Scheme SSI}
In this scenario, schematically shown in Fig.~\ref{fig:3l2h},   
it holds $m_5> m_4> m_2> m_1> m_3$, i.e., 
the two heavy sterile neutrinos are heavier than the three 
light neutrinos which enjoy an inverted hierarchy. 
Consequently, Eq.~(\ref{eq:3IO}) holds. In addition, we have 
\be
\lsndo = m_5^2 - m_3^2 ~~\mbox{ and } ~~\lsndt = m_4^2 - m_3^2~,
\ee
and the masses in terms 
of the smallest mass are 
\bea \label{eq:54Imass}
m_1 = \sqrt{\dma + m_3^2}\,,~m_2 = \sqrt{\dma + \dms + m_3^2}~,\\[0.2cm]
m_4 = \sqrt{\lsndt + m_3^2}\,,~m_5 = \sqrt{\lsndo + m_3^2}~.
\eea
Neglecting the smallest mass we find typical values of 
$m_2 \simeq m_1 \simeq \sqrt{\dma} \simeq 0.05$ eV, 
$m_4 \simeq \sqrt{\lsndt} \simeq 0.94~(0.95)$ eV and 
$m_5 \simeq \sqrt{\lsndo} \simeq 2.55~(1.38)$ eV. 
The active neutrinos behave according to an inverted hierarchy 
in a 3-generation framework. As a function 
of the smallest 
neutrino mass $m_3$ the other masses are shown in Fig.~\ref{fig:masses}. 
One finds that $\Sigma^{\rm SSI}$ 
is identical to $\Sigma^{\rm SSN}$ in Eq.~(\ref{eq:Sigma54N}) 
(\dms~has to be replaced with \dma, which does not make a notable 
difference). 
The kinematic mass is also basically identical to the one in scenario 
SSN, which is given in Eq.~(\ref{eq:KATRIN54N}).   
In the effective mass the situation is different\footnote{There 
is another non-oscillation probe which can distinguish SSN and SSI, 
namely the decay of astrophysical high energy neutrinos, treated 
in Appendix \ref{sec:UHE}.}, because the light 
neutrinos obey an inverted hierarchy and therefore a contribution of 
the same order of magnitude as the sterile ones: 
\bea \label{eq:meff54I}
\meff^{\rm SSI} \simeq 
\left| \sss \, \sqrt{\dma} + \css \, \sqrt{\dma} \, 
e^{i \alpha_2} \right. \\[0.2cm] 
\left. + |U_{e4}|^2 \, \sqrt{\lsndt} \, e^{i \alpha_4} 
+ |U_{e5}|^2 \, \sqrt{\lsndo} \, e^{i \alpha_5 } \right| ~.
\eea 
The absolute value of the first two terms is (see 
Table \ref{tab:3mass}) 
$\sqrt{\dma} \, \sqrt{1 - \sin^2 2 \theta_\odot \, \sin^2 \alpha_2 /2}$ 
(between 0.020 and 0.051 eV) 
while for the best-fit values from Eq.~(\ref{eq:stBF}) 
the absolute value of 
the last two terms is between 0.025 and 0.048 eV,  
see Eq.~(\ref{eq:meff54N}). Hence, 
the effective mass can vanish completely in this scheme even for 
the best-fit values and a vanishing smallest neutrino 
mass. This is borne out by the dark shaded regions in  
the upper right panels of  
Figs.~\ref{fig:meff1} and \ref{fig:meff2} for the best-fit points of 
Eqs.~(\ref{eq:stBF}) and (\ref{eq:st2}). 
The effective mass can be as large as 0.1 eV (or 0.08 eV for the second 
typical point from Eq.~(\ref{eq:st2})) in the small $m_1$ ($\ls 0.1$ eV) 
regime. Finding a larger \meff~will rule out scenario SSI. 

We remark here that only in the schemes SSN and SSI 
the magnitudes of $U_{e4}$ and 
$U_{e5}$ are important for the predictions of $m_\beta$ and \meff. However, 
our statements regarding the possible exclusion of scenarios  
SSN and SSI with future measurements is rather insensitive to the 
precise values of  $U_{e4}$ and $U_{e5}$.

\section{\label{sec:3h2l}Sterile Neutrinos lighter than 
active Neutrinos: 3+2 Scenarios}

\subsection{\label{sec:N45}Scheme NSS}
In this scheme, $m_3> m_2> m_1> m_4> m_5$, i.e., the 
three normally ordered active neutrinos are heavier than the two 
sterile neutrinos, see Fig.~\ref{fig:3h2l}. 
Apart from Eq.~(\ref{eq:3NO}) it holds 
that 
\be
\lsndo = m_3^2 - m_5^2 ~~\mbox{ and }~~ \lsndt = m_3^2 - m_4^2~,
\ee
and the masses in terms of the smallest mass and the mass-squared differences 
are 
\bea \label{eq:N45mass}
m_4 = \sqrt{\lsndo - \lsndt + m_5^2}\,,~
m_1 = \sqrt{\lsndo - \dma - \dms + m_5^2}~,\\[0.2cm]
m_2 = \sqrt{ \lsndo - \dma + m_5^2}\,,~m_3 = \sqrt{\lsndo + m_5^2}~.
\eea
We therefore have for a negligible smallest mass 
three quasi-degenerate neutrinos and another quite massive state. 
Their values are  
$m_{1} \simeq m_2 \simeq m_3 \simeq \sqrt{\lsndo} \simeq 2.55~(1.38)$ 
eV and $m_4 \simeq \sqrt{\lsndo - \lsndt} \simeq 2.37~(1.00)$ eV. 
As a function 
of the smallest mass $m_5$, they are given in Fig.~\ref{fig:masses}.

Let us neglect the smallest neutrino mass and insert the best-fit 
values. In this case, the sum of masses as constrained by 
cosmological observations is given by  
\be \label{eq:SigmaN45}
\Sigma^{\rm NSS} \simeq  \sqrt{\lsndo - \lsndt} + 3 \sqrt{\lsndo} 
\simeq 
10.0~(5.14)~\rm eV ~.
\ee
The kinematic mass is given by 
\bea \label{eq:KATRINN45}
m_\beta^{\rm NSS} \simeq 
\sqrt{\lsndo \, (1 - |U_{e5}|^2)}  
\simeq 
2.55 ~(1.38)~\rm eV~. 
\eea
With $|U_{e5}|^2$ being very small there is basically no dependence on the 
mixing parameters $|U_{ei}|^2$ 
because of the quasi-degenerateness of the three leading neutrinos. 
Only for $m_5$ reaching eV values one observes deviations from the 
last two equations in Fig.~\ref{fig:mass1}.
The cosmological observable $\Sigma$ is quite large, namely 
between 9 and 11 (3 and 6) eV. Interestingly, 
$m_\beta \simeq \sqrt{\lsndo}$ lies always 
above 1 eV, and is even above the current bound of 
2.3 eV for the best-fit point. Hence, this scenario 
constraints $\lsndo$ to lie below $\simeq 5.3$ eV$^2$. 
In general, if this scenario is realized, KATRIN will definitely 
observe a positive signal the absence of which 
can rule out this scenario.

Finally, for \onbb~we have (in the limit $|U_{e4}| \rightarrow 0$)  
\bea \label{eq:meffN45}
\meff^{\rm NSS} \simeq 
\sqrt{\lsndo} \, \sqrt{1 - \sin^2 2 \theta_\odot 
\, \sin^2 \alpha_2/2 } 
\simeq  
(1.02 \div 2.55)~(0.55 \div 1.38)\rm~eV ~,
\eea 
which is dominated by the three active quasi-degenerate neutrinos and cannot 
vanish due to the non-maximality of solar neutrino mixing  
as is reflected in the lower left panel of Fig.~\ref{fig:meff1}.
The effective mass ranges from $\sqrt{\lsndo} \, \cos 2 \theta_\odot$ 
to $\sqrt{\lsndo}$. 
In general, the effective mass is sizable in scenario NSS
and already a part of it around the best-fit point 
is disfavored by the current upper limit of 1 eV. 
In fact, 
improving the limit on the effective mass below 0.2 eV rules out this 
scheme if neutrinos are Majorana particles.

It is possible to 
set limits on \lsndo~and in particular 
on $\theta_\odot$ and the Majorana phase $\alpha_2$  
demanding \meff~to lie within a specific limit.
Using the 3$\sigma$ ranges from 
Eq.~(\ref{eq:data}) and the ranges around the best-fit point 
from Eq.~(\ref{eq:stBF}) one can investigate what values are allowed. 
The result for $\alpha_2$ and $\sss$         
can be seen in Fig.~\ref{fig:meff_exa}. We took for \meff~the 
current limit of 1 eV and a future limit of 0.5 eV. Taking the second central 
point for the sterile neutrino parameters gives hardly any constraint for 
a limit of 1 eV, but for $\meff \le 0.5$ eV the plot looks similar to 
the 1 eV plot of the best-fit point.

\subsection{\label{sec:N54}Scheme ISS}
In this scheme (see Fig.~\ref{fig:3h2l}) 
it holds $m_2> m_1> m_3> m_4> m_5$, i.e., the 
three inversely ordered active neutrinos are heavier than the two 
sterile neutrinos. Apart from Eq.~(\ref{eq:3IO}) we have  
\be
\lsndo = m_2^2 - m_5^2 ~~\mbox{ and }~~ \lsndt = m_2^2 - m_4^2~,
\ee
and the masses in terms of the smallest mass are 
\bea \label{eq:N54mass}
m_4 = \sqrt{\lsndo - \lsndt + m_5^2}\,,~
m_3 = \sqrt{\lsndo - \dma - \dms + m_5^2}~,\\[0.2cm]
m_1 = \sqrt{ \lsndo - \dms + m_5^2}\,,~m_2 = \sqrt{\lsndo + m_5^2}~.
\eea
The results are basically identical to scenario NSS. 
We therefore do not give the expressions. Distinguishing between scenarios 
ISS and NSS could for instance be done via matter effects in oscillation 
experiments or in supernovae \cite{3+2phen0}.

\section{\label{sec:131}One heavy and one light 
sterile Neutrino: 1+3+1 Scenarios}

We will discuss now the mass-related observables when the three active 
neutrinos are `''sandwiched'' between the sterile ones. Recall that 
the fit from \cite{3+2data} does not apply in this case.  We note that 
apart from the mass-squared differences the 
mixing matrix elements $U_{e4}$ and $U_{e5}$ might also be different. 
However, their values do hardly influence the predictions. 
We will insert for 
the rest of this Section the numerical values from Eqs.~(\ref{eq:stBF}, 
\ref{eq:st2}) for the sterile 
neutrino parameters, but will indicate that there might be differences by 
replacing in the expressions 
$\lsndo \rightarrow \lsndot$ and $\lsndt \rightarrow \lsndtt$.

\subsection{\label{sec:5N4}Scheme SNSa}
In this scheme, $m_5> m_3> m_2> m_1> m_4$. 
Apart from Eq.~(\ref{eq:3NO}) 
we have 
\be
\lsndot = m_5^2 - m_1^2 ~~\mbox{ and }~~ \lsndtt = m_1^2 - m_4^2~, 
\ee
see Fig.~\ref{fig:131}. We can express the individual masses as 
\bea \label{eq:5N4mass}
m_1 = \sqrt{\lsndtt + m_4^2}\,
,~m_2 = \sqrt{\lsndtt + \dms + m_4^2}~,\\[0.2cm]
m_3 = \sqrt{\lsndtt + \dms + \dma + m_4^2}\,,~
m_5 = \sqrt{\lsndot + \lsndtt + m_4^2}~.
\eea
We therefore have three quasi-degenerate active neutrino masses of order 
$\sqrt{\lsndtt} \simeq 0.94~(0.95)$ eV and one very heavy mass around 
$\sqrt{\lsndot + \lsndtt} \simeq 2.71~(1.67)$ eV. 
A plot of the $m_i$ as a function of $m_4$ is 
given in Fig.~\ref{fig:masses}. Neglecting the smallest mass is 
correct as long as it is below 0.5 eV. 
We can estimate that 
\bea \label{eq:Sigma5N4}
\Sigma^{\rm SNSa} \simeq 
3 \sqrt{\lsndtt} + \sqrt{\lsndot + \lsndtt} \simeq 5.55~(4.52)~\rm eV ~.
\eea
The magnitude of $\Sigma$ is quite sizable,  
and can be 
between 5 and 6 (4 and 5) eV, 
if one varies the mass-squared differences 
in their allowed ranges.  

The sterile masses provide the leading contribution 
not only in this observable, 
but also in the kinematic mass: 
\bea \label{eq:KATRIN5N4}
m_\beta^{\rm SNSa} \simeq \sqrt{\lsndtt + |U_{e5}|^2 \, 
(\lsndot + \lsndtt)} \simeq 
1.00~(0.97)~\rm eV ~.
\eea 
As in scenarios NSS and ISS, $m_\beta \simeq \sqrt{\lsndtt}$ 
is always above the KATRIN sensitivity 
of 0.3 eV, therefore a signal in this experiment corresponding to 
at least 1 eV should be observed if this scheme is realized. 

Finally, \onbb~should be triggered by an effective mass given by  
\bea \label{eq:meff5N4}
\meff^{\rm SNSa} \simeq 
\left| \css \, \sqrt{\lsndtt} + \sss \, \sqrt{\lsndtt} \, 
e^{i \alpha_2} \right. \\[0.2cm]
\left. + |U_{e5}|^2 \, \sqrt{\lsndot + \lsndtt} \, e^{i \alpha_5} 
\right| \simeq \sqrt{\lsndtt} \, 
\sqrt{1 - \sin^2 2 \theta_\odot \, \sin^2 \alpha_2/2 } ~,
\eea 
where the third term can be neglected. Hence, 
$\meff^{\rm SNSa}$ is dominated by the active neutrinos and 
lies between $\sqrt{\lsndtt}$ 
and $\sqrt{\lsndtt} \, \cos 2 \theta_\odot$, which is roughly 
0.9 and 0.38 eV, respectively. 
The effective mass for this scenario is plotted in the lower right panel of 
Fig.~\ref{fig:meff1} for the best-fit values 
as well as by varying all parameters within  
their allowed range.   
The situation for \meff~is unfortunately similar to scenarios NSS and ISS, 
even though here the overall mass scale is $\sqrt{\lsndtt}$, while 
it was $\sqrt{\lsndo}$ in the previous cases. The problem is of course 
the allowed range of the mass-squared differences and the unknown 
Majorana phase. Only if the condition 
\be \label{eq:cond}
\sqrt{\lsndo} \, \cos 2 \theta_\odot \ge  \zeta~\sqrt{\lsndtt} 
\ee
is fulfilled, then we can distinguish scenarios NSS/ISS and SNSa/SISa 
via \onbb. 
In Eq.~(\ref{eq:cond}) we have included a factor $\zeta \ge 1$, 
which takes into account the nuclear matrix element uncertainty, 
a necessity when one tries to distinguish different mass orderings 
via \onbb~\cite{meff_spec}. 
For the best-fit values and $\zeta=1$ indeed Eq.~(\ref{eq:cond}) 
is fulfilled, but already 
for the second central point one cannot distinguish the schemes anymore 
as can also be seen from the lower panels of Fig.~\ref{fig:meff2}. 

Anyway, if neutrinos are Majorana particles, then we can rule out 
scenario SNSa if $\meff \le 0.1$ eV. 
One can generate plots as 
shown in Fig.~\ref{fig:meff_exa} in order to obtain constraints 
on the parameters 
$\theta_\odot$ and $\sin^2 \alpha_2$ from experimental 
information about \meff. This requires limits which are stronger 
by a factor $\sqrt{\lsndo/\lsndtt}$ than the 
limits used to generate Fig.~\ref{fig:meff_exa}. 

\subsection{\label{sec:4N5}Scheme SISa}
In this scheme we have $m_5> m_2> m_1> m_3> m_4$, i.e., the sterile 
neutrinos are above and below three inversely ordered active neutrinos, 
see Fig.~\ref{fig:131}.
One finds 
\be
\lsndot = m_5^2 - m_3^2 ~~\mbox{ and } ~~\lsndtt = m_3^2 - m_4^2~,
\ee
and can express the individual masses as 
\bea \label{eq:5I4mass}
m_3 = \sqrt{\lsndtt + m_4^2}\,,~
m_1 = \sqrt{\lsndtt + \dma + m_4^2}~,\\[0.2cm]
m_2 = \sqrt{\lsndtt + \dma + \dms + m_4^2}\,,~
m_5 = \sqrt{\lsndot + \lsndtt + m_4^2}~.
\eea
We do not give the expressions for $\Sigma$, $m_\beta$ or \meff, because 
the results are indistinguishable from scenario SNSa. Again, 
matter effects in neutrino oscillation experiments could be used to 
distinguishing the scenarios. 

\subsection{\label{sec:5N4b}Scheme SNSb}
In this scheme, $m_4> m_3> m_2> m_1> m_5$. 
Apart from Eq.~(\ref{eq:3NO}) we have 
\be
\lsndot = m_1^2 - m_5^2 ~~\mbox{ and }~~ \lsndtt = m_4^2 - m_1^2~, 
\ee
see Fig.~\ref{fig:131}. The individual masses are  
\bea \label{eq:SNSbmass}
m_1 = \sqrt{\lsndot + m_5^2}\,,~
m_2 = \sqrt{\lsndot + \dms + m_5^2}~,\\[0.2cm]
m_3 = \sqrt{\lsndot + \dms + \dma + m_5^2}\,,~
m_4 = \sqrt{\lsndot + \lsndtt + m_5^2}~.
\eea
The mass-related observables are obtained from the formulae for 
scenario SNSa from Section \ref{sec:5N4} with the exchange 
$\lsndot \leftrightarrow \lsndtt$. Hence, 
\bea \label{eq:Sigma5N4b}
\Sigma^{\rm SNSb} \simeq 
3 \sqrt{\lsndot} + \sqrt{\lsndot + \lsndtt} 
\simeq 10.36~(5.81)~\rm eV ~.
\eea
\bea \label{eq:KATRIN5N4b}
m_\beta^{\rm SNSb} 
\simeq \sqrt{\lsndot + |U_{e5}|^2 \, (\lsndot + \lsndtt)} 
\simeq 
2.57~(1.39)~\rm eV ~.
\eea 
\bea \label{eq:meff5N4b}
\meff^{\rm SNSb} \simeq 
\left| \css \, \sqrt{\lsndot} + \sss \, \sqrt{\lsndot} \, 
e^{i \alpha_2} \right. \\[0.2cm]
\left. + |U_{e5}|^2 \, \sqrt{\lsndot + \lsndtt} \, e^{i \alpha_5} 
\right| \simeq \sqrt{\lsndot} \, 
\sqrt{1 - \sin^2 2 \theta_\odot \, \sin^2 \alpha_2/2 } ~.
\eea 
All these expression are almost identical to the ones 
for scenarios NSS and ISS, because the leading contributions to all 
observables correspond to a situation with three quasi-degenerate 
active neutrinos having a mass $\sqrt{\lsndot}$. 
Therefore mass-related observables can  
not distinguish these cases, unless the mass-squared 
differences \lsndo~and \lsndot~are very much different from each other. 

\subsection{\label{sec:4N5b}Scheme SISb}
In this scheme we have $m_4> m_2> m_1> m_3> m_5$, i.e., the sterile 
neutrinos are above and below three inversely ordered active neutrinos, 
see Fig.~\ref{fig:131}.
One finds 
\be
\lsndot = m_3^2 - m_5^2 ~~\mbox{ and } ~~\lsndtt = m_4^2 - m_3^2~,
\ee
and can express the individual masses as 
\bea \label{eq:5I4bmass}
m_3 = \sqrt{\lsndot + m_5^2}\,,~
m_1 = \sqrt{\lsndot + \dma + m_5^2}~,\\[0.2cm]
m_2 = \sqrt{\lsndot + \dma + \dms + m_5^2}\,,~
m_4 = \sqrt{\lsndot + \lsndtt + m_5^2}~.
\eea
We do not give the expressions for $\Sigma$, $m_\beta$ or \meff, because 
the results are indistinguishable from scenario SNSb and 
therefore also from NSS and ISS. Again, 
matter effects in neutrino oscillation experiments could be used to 
distinguish between the scenarios.

\section{\label{sec:concl}Discussions and Summary}
Adding two sterile neutrinos to the three active ones 
gives rise to eight possible mass 
orderings, out of which the right one should be identified 
in order to pin down the flavor structure 
of the neutrino mass matrix. We have investigated how and if mass-related  
measurements can do the job. 
In addition, we studied 
the general properties of the non-oscillation observables in scenarios 
with two sterile neutrinos. 
The possible mass orderings are 
shown schematically in Figs.~\ref{fig:3l2h}, \ref{fig:3h2l} and 
\ref{fig:131}. 
Apart from the usual 3-generation masses and mixing 
parameters 
we have to cope with two additional mixing matrix elements 
$|U_{e4}|$ and $|U_{e5}|$ as well as with 
two mass-squared differences $\lsndo$ and $\lsndt$.   
Without loss of generality we can assume $\lsndo > \lsndt$ and associate   
$\lsndo$ with the state 5 and $\lsndt$ with state 4, respectively.  

We use the following nomenclature for the eight different schemes:  
\begin{itemize} 
\item[(i)] SSX, where X = N for a normal and 
X = I for an inverted ordering of the mostly active neutrinos. 
In these schemes the two
 predominantly sterile neutrinos are heavier 
than the three predominantly active neutrinos (2+3 scenarios); 
\item[(ii)] XSS (X = N or I as before), where the two 
 predominantly sterile neutrinos are lighter than the three 
predominantly active neutrinos (3+2 scenarios); 
\item[(iii)] SXS with X = N or I, where the three active neutrinos 
are sandwiched between the sterile ones (1+3+1 scenarios). In 
this class there can be four possible 
scenarios which we denote as SXSa and SXSb. 
The scheme SXSa corresponds to the state 5 higher 
than the three active states 
and SXSb corresponds to the state 
5 lower than the three active states. 
Those scenarios are strictly speaking not covered by the available 
analyzes of scenarios with two sterile neutrinos. In absence of 
any fit of this possibility, we assumed for simplicity that the 
parameters are the same as for the other scenarios. 
\end{itemize}

The following general comments can be made about the 
different mass related observables\footnote{Predictions 
for mass-related observables in the 
presence of yet another sterile neutrino are discussed in 
Appendix \ref{sec:3+3}.}: The sum of neutrino masses 
depends basically only on the new mass-squared differences, and 
typical (minimal) values are 
$\sqrt{\lsndo} + \sqrt{\lsndt}$, $3\sqrt{\lsndo}$ 
or $3\sqrt{\lsndt}$,  
depending on the mass ordering. 
Given the best-fit values and allowed ranges of masses this is already 
in conflict with the standard cosmological scenario, as discussed below.

\begin{table}[t]
\begin{center}
\begin{tabular}{|c|c|c|c|}\hline
 & $m_{\beta}$ (eV) & $\meff$ (eV) & $\Sigma$ (eV) \\ \hline \hline
SSN & 0.15 $\div$ 0.52 & 0.0 $\div$ 0.11 & 3.33$\div$ 3.85  \\ \hline
SSI & 0.16 $\div$ 0.52  & 0.0 $\div$ 0.16 & 3.33 $\div$ 3.85  \\ \hline
NSS, ISS, SNSb, SISb & 2.3  $\div$ 2.7 & 0.43 $\div$ 2.72 
& 9.16 $\div$ 10.80  \\ \hline
SNSa, SISa & 0.89 $\div$ 1.11 & 0.09 $\div$ 1.03 & 5.19 $\div$ 5.92 \\ \hline
Current Bound & 2.2 eV (95\% C.L.)  & $\sim$1 eV (90\% C.L.) & $\sim$1 eV \\  \hline
\end{tabular}
\caption{\label{tab:range}The ranges of the 
predictions for $m_\beta$, $\meff$ and $\Sigma$ according to 
the various mass orderings for $m_{\mathrm{smallest}}< 0.1$ eV and 
assuming that $\lsndo = \lsndot$ and $\lsndt = \lsndtt$. 
Also shown are the current bounds on these observables. 
}
\end{center} 
\end{table}

The parameters relevant for \onbb~and direct beta-decay searches 
can be written as a contribution from the three mostly active states 
and the two mostly sterile neutrinos: 
\be \label{eq:3+st}
\meff = \left| \meff^{3} + \meff^{\rm st} \right| ~\mbox{ and } ~
m_\beta = \sqrt{(m_\beta^{3})^2 + (m_\beta^{\rm st})^2}~.
\ee
where $|\meff^{3}|$ and $m_\beta^{3}$ are the expressions known from 
3-flavor analyzes, see Table \ref{tab:3mass}. 
All cases are possible in Eq.~(\ref{eq:3+st}): 
dominance of the sterile contribution, dominance of the active
contribution, and equal-sized contributions, leading (only in \meff) 
potentially to complete cancellation. 

In general, the mass-related observables
can not distinguish between a normal or inverted ordering 
of the three active neutrinos, with the exception of schemes 
SSN and SSI, which have different predictions for \meff.  
It will however be difficult to test this difference in practise, as 
precise knowledge of the oscillations parameters is required. 
Only in the schemes SSN and SSI the magnitude of 
$|U_{e4}|$ and $|U_{e5}|$ is crucial for the predictions of 
\meff~and $m_\beta$. 
For all other mass orderings 
the dependence on $|U_{e4}|$ and $|U_{e5}|$ is suppressed 
($\Sigma$ does not depend on $|U_{e4}|$ and $|U_{e5}|$). 
Scenarios SNSb and SISb 
are indistinguishable from scenarios NSS and ISS if the mass-squared 
differences are equal or very similar. 
It turns out that in order to summarize all phenomenology 
of the mass-related observables it suffices to plot them 
for four schemes: SSN, SSI, SNSa (covering also SISa) and 
NSS (covering also ISS, SNSb, SISb). Interestingly, these four 
cases have also the same phenomenology in what regards decays of 
high energy astrophysical neutrinos, see Appendix \ref{sec:UHE}. 
In Table \ref{tab:range} we present 
for the parameter ranges given in 
Eqs.~(\ref{eq:data}) and (\ref{eq:stBF}) 
the predictions of the three quantities 
$m_\beta$, $\meff$, $\Sigma$ for the four types of mass orderings in 
the realistic case when the smallest neutrino mass is 
smaller than 0.1 eV. Also shown are the current bounds on these observables.
Table \ref{tab:scheme} summarizes what the various schemes mean for 
KATRIN and for future experiments searching for \onbb. 
The following conclusions can be drawn:

\begin{table}[ht]
\begin{center}
\begin{tabular}{|c|c|c|c|}\hline
scheme & feature &  KATRIN & \obb  \\ \hline \hline
SSN & NH plus $\nu_{s_1}$, $\nu_{s_2}$ & maybe & maybe  \\ \hline
SSI & IH plus $\nu_{s_1}$, $\nu_{s_2}$ & maybe & maybe \\ \hline
NSS, ISS & QD with $\sqrt{\lsndo}$ &  yes & yes  \\ \hline
SNSb, SISb & QD with $\sqrt{\lsndot}$ & yes & yes  \\ \hline
SNSa, SISa & QD with $\sqrt{\lsndtt}$ & yes & yes  \\ \hline
\end{tabular}
\caption{\label{tab:scheme}The various schemes with two sterile neutrinos 
and their meaning for KATRIN and future \obb~experiments. We assumed 
that $\lsndot$ and $\lsndtt$ are larger than 0.1 eV$^2$. }
\end{center}
\end{table}
\begin{itemize}
\item scenarios SSN and SSI predict for all observables the smallest 
values. Hence, they are the easiest to rule out. 
The other scenarios 
correspond at leading order to quasi-degenerate 3-neutrino scenarios 
with the common mass scale given by $\sqrt{\lsndo}$ (NSS, ISS), 
$\sqrt{\lsndtt}$ (SNSa, SISa) or $\sqrt{\lsndot}$ (SNSb, SISb); 
\item model independent constraints on neutrino masses stem from 
direct searches in the spectra of beta-decays. 
The Mainz data give the constraint $m_\beta < 2.2$ eV at 
95\% C.L. ($\Delta \chi^2 =4$) \cite{lisi}. Following the procedure in 
\cite{lisi} the 68\% C.L.~($\Delta \chi^2 =1$) bound is 0.7 eV and the 
99.73\% C.L.~($\Delta \chi^2 =9$) bound is 4.0 eV. 
Scenarios 
SSN and SSI with both 
sterile neutrinos being heavier than the active ones can 
have unobservably small $m_\beta$ and are consistent with the 
current bound even at 1$\sigma$ as can be seen from Table 2. 
The Table also shows that scenarios SNSa and SISa are allowed 
at 2$\sigma$, while the scenarios NSS/ISS/SNSb/SISb are consistent with 
the Mainz result at 3$\sigma$. 
The other six mass orderings 
will definitely result in a signal in KATRIN. If the two sterile 
neutrinos are lighter than the active ones (NSS and ISS), then there is 
a direct correspondence $m_\beta \simeq \sqrt{\lsndo}$, which can be used 
to rule out part of the allowed range of $\lsndo$ already at the current 
stage. Scenarios SNSa and SISa predict 
$m_\beta \simeq \sqrt{\lsndtt}$, which will rule out part of its 
allowed range in the near future; 

\item 
the effective mass governing \onbb~can only vanish when the sterile 
neutrinos are heavier than the active ones 
(SSN and SSI, for which the largest value is 0.1 eV). In 
the other orderings the non-maximality of the solar neutrino mixing angle 
renders \meff~non-zero. 
If the sterile neutrinos are lighter than the active 
ones (NSS and ISS), then 
\meff~is larger than 0.2 eV and has a maximal value above 
the current limit\footnote{Note that 
in our analysis we are not using the data coming from the positive 
evidence claimed by a part of the Heidelberg-Moscow collaboration 
\cite{klapdor} and thus we have only an upper bound.} 
of $\sim$ 1 eV at 90\% C.L.
Consequently, this scenario can be ruled out by a stronger limit 
on \meff~and one can also constrain parameters 
with the current limit.  This 
concerns in particular $\sss$ and the Majorana phase $\alpha_2$. 
Unfortunately, the sterile neutrino parameters are such that 
this scenario is 
hardly distinguishable from the scenarios 
in which one sterile neutrino 
is heavier and the other one lighter than the active ones. 
Whereas telling apart NSS/ISS from SNSb/SISb is basically impossible if 
$\lsndo$ is similar to $\lsndot$, 
distinguishing SNSa/SISa from NSS/ISS requires a condition of the 
form 
$\sqrt{\lsndo} \, \cos 2 \theta_\odot \ge  \zeta~\sqrt{\lsndtt}$, 
where $\zeta$ denotes the nuclear matrix element uncertainty.  
\end{itemize}

In general, the sum of neutrino masses in the scenarios under study 
is always larger than about 2 eV. 
This minimal 
value is obtained in the two schemes SSN and SSI in which the 
sterile neutrinos are heavier than the active ones. In addition, the 
two mass-squared differences related to the MiniBooNE/LSND experiments 
should be rather small, because $\Sigma = \sqrt{\lsndo} + \sqrt{\lsndt}$. 
The other scenarios have sizable values of $\Sigma$, approaching up to 
10 eV if the sterile neutrinos are lighter than the active ones.  
However, cosmological limits can always be evaded or relaxed. 
Nevertheless, we add some discussion on typical limits obtained 
in the literature (for an overview, see \cite{steen}), 
which typically constrain both the sum of 
neutrino masses and the effective 
number of neutrino species $N_{\rm eff}$ 
contributing to the radiation density. Those limits 
depend also on $N_{\rm m}$, which is the number of
{\it equally massive} species. 
As one example, we focus on Ref.~\cite{raffelt06}, in which 
likelihood contours are 
provided in the $N_{\rm eff}$--$\Sigma$ plane for three cases:  
(i) $N_{\rm m} = N_{\rm eff}$,  
~(ii) $N_{\rm m} = 3$ and 
(iii) $N_{\rm m} = 1$. The value 
$N_{\rm eff} = 5$ is allowed only at about 99\% C.L.~in all the above cases
and the bounds on $\Sigma$ are 0.62, 0.57 and 0.41 eV, respectively 
(all at 95\% C.L.). The type Ia supernova data from SNLS, 
large scale structure data from 2DF and SDSS, baryon acoustic oscillation 
data from SDSS, CMB anisotropy data from 
WMAP and the smaller scale measurement by the BOOMERANG experiment were 
included in that analysis. Adding the Lyman-$\alpha$ forest data gives even 
stronger bounds \cite{lyman}, leaving out the baryon acoustic 
oscillation relaxes 
the limits \cite{steen}. 
For the unrealistic case when the lightest neutrino is heavier than 
1 eV we have $N_{\rm m} = N_{\rm eff} = 5$ and case (i) applies. The 
resulting $\Sigma$ in our scenarios is of course much larger than allowed. 
Another example is when in scenarios NSS/ISS the smallest mass can be neglected. 
Then we have three quasi-degenerate neutrinos, but also one other massive 
neutrino with mass $\sqrt{\lsndo - \lsndt}$. This possibility, 
as well as the other cases we study,  
is not covered by the analysis in \cite{raffelt06} or in any other 
paper we are aware of. 
Nevertheless, we can safely assume\footnote{
A more quantitative and accurate estimate on the
joint constraint on  the number of neutrino species and the sum of neutrino
masses from cosmology in the various scenarios
 and estimating the $\Delta\chi^2$ would
require a more detailed analysis of the cosmological data sets which
is clearly not in the purview of the present analysis.} 
that limits on the sum of masses for 
$N_{\rm eff} = 5$ do not exceed $\sim 1$ eV.
Consequently, and not surprisingly, 
all scenarios with two sterile neutrinos have serious problems 
with cosmology and require non-standard physics 
(primordial lepton asymmetries, low reheating temperature, 
additional neutrino interactions,$\ldots$) as described, e.g., 
in \cite{cosmo2,cosmo}.

To conclude, scenarios with two sterile neutrinos 
offer rich and interesting phenomenology. In six of the eight allowed cases  
KATRIN and future \obb~experiments will find a signal. 
Mass-related observables alone, however, can not identify 
the correct mass ordering completely, which leaves room for 
further studies in order to disentangle the possibilities by means 
of oscillation experiments.

\vspace{0.5cm}
\begin{center}
{\bf Acknowledgments}
\end{center}
We thank T.~Schwetz for discussions and 
providing us with numerical results.  
This work was supported by the Alexander-von-Humboldt-Foundation (S.G.). 
W.R.~acknowledges support 
by the ``Deutsche Forschungsgemeinschaft'' 
in the Transregio 27 ``Neutrinos and beyond -- Weakly 
interacting particles in Physics, Astrophysics and Cosmology'' 
and under project number RO--2516/3--2, as well as by the  
EU program ILIAS N6 ENTApP WP1. 
S.G.~wishes to thank the Max--Planck--Institut f\"ur Kernphysik, 
Heidelberg, for hospitality. 

\begin{appendix}

\renewcommand{\theequation}{A\arabic{equation}}
\setcounter{equation}{0}

\section{\label{sec:3+3}On Scenarios with three sterile Neutrinos}

For the sake of completeness we summarize briefly the formulae and 
results for the mass-related observables in case when three sterile 
neutrinos are added. 
No qualitatively new aspects are found in these scenarios. 
The authors of Ref.~\cite{3+2data} also performed an analysis of 
this possibility and it was found that no significant improvement 
of the fit can be achieved in this way. 
The best-fit values for the mass-squared differences are 
\be \label{eq:33BF}
\lsndo = 1.84~{\rm eV}\,,~\lsndt = 0.83~{\rm eV}\,,
~\lsndth = 0.46~{\rm eV}~.
\ee
Note that there can be scenarios in which the fit of Ref.~\cite{3+2data} 
does not apply. 
There is no information given on the mixing matrix elements, 
let us therefore take for simplicity the values 
\be
|U_{e i}| = 0.1~\mbox{ for } i = 4,5,6~.
\ee
We will estimate now the values of the mass-related observables for 
all possible mass orderings. Again, we fix $m_{1,2,3}$ to be 
responsible for the oscillations of solar and atmospheric 
neutrino oscillations. 
With the details given in the main text, 
it is quite easy to obtain the following formulae, which are valid when 
the smallest neutrino mass is neglected. As for two sterile 
neutrinos, the non-oscillation probes can not distinguish whether the 
active neutrinos are normally or inversely ordered. There 
are in total 16 possible mass orderings, which we group in 4 classes: 

\begin{itemize}
\item[(i)]
the first type of mass spectrum holds when the three sterile neutrinos 
are all heavier than the active ones (SSSN and SSSI). 
In this case, 
\bea 
\Sigma \simeq \sqrt{\lsndo} + \sqrt{\lsndt} + \sqrt{\lsndth} 
\simeq 2.9~{\rm eV}~,\\[0.2cm]
m_\beta \simeq \sqrt{ |U_{e 4}|^2 \, \lsndth + 
|U_{e 5}|^2 \, \lsndt + |U_{e 6}|^2 \, \lsndo } 
\simeq 0.18~{\rm eV}~,\\[0.2cm]
\meff \simeq \left| |U_{e4}|^2 \, \sqrt{\lsndth} + 
|U_{e5}|^2 \, \sqrt{\lsndt} \, e^{i\alpha_{54}} 
+ |U_{e6}|^2 \, \sqrt{\lsndo} \, e^{i\alpha_{64}}
\right| \ls 0.29~{\rm eV}~,
\eea
where $\alpha_{54} = \alpha_5 - \alpha_4$ 
and $\alpha_{64} = \alpha_6 - \alpha_4$ are combinations of Majorana 
phases. The situation is somewhat similar to scenarios SSN and SSI; 
\item[(ii)]a second class of spectra is found when the 
three sterile neutrinos are lighter than the three active 
ones (scenarios NSSS and ISSS). One has 
\bea 
\Sigma \simeq 3 \sqrt{\lsndo} \simeq 4.1~{\rm eV}~,\\[0.2cm]
m_\beta \simeq \sqrt{ \lsndo } \simeq 1.36~{\rm eV}~,\\[0.2cm]
\meff \simeq \sqrt{ \lsndo } \, 
\sqrt{1 - \sin^2 2 \theta_\odot \, \sin^2 \alpha_2/2} 
\simeq (0.54 \div 1.36)~{\rm eV}~.
\eea
This is similar to the possibilities NSS and ISS; 
\item[(iii)]two neutrinos can be heavier than the 
active ones which in turn are 
heavier than the third sterile state. 
There are three possibilities for this.  
If the two heavy sterile 
neutrinos correspond to \lsndo~and \lsndt, then 
the resulting schemes SSNSa and SSISa give
\bea 
\Sigma \simeq 3 \sqrt{\lsndth} + 
\sqrt{\lsndth + \lsndt} + \sqrt{\lsndth + \lsndo} 
\simeq 4.7~{\rm eV}~,\\[0.2cm]
m_\beta \simeq \sqrt{ \lsndth + |U_{e5}|^2 \, (\lsndth + \lsndt) 
+ |U_{e6}|^2 \, (\lsndth + \lsndo) } \simeq 0.70~{\rm eV}~,\\[0.2cm]
\meff \simeq \sqrt{ \lsndth } \, 
\sqrt{1 - \sin^2 2 \theta_\odot \, \sin^2 \alpha_2/2} 
\simeq (0.27 \div 0.68)~{\rm eV}~.
\eea
Also possible is that the two heavy sterile 
neutrinos correspond to \lsndo~and \lsndth~
(schemes SSNSb and SSISb), in which case 
\bea 
\Sigma \simeq 3 \sqrt{\lsndt} + 
\sqrt{\lsndt + \lsndth} + \sqrt{\lsndt + \lsndo} 
\simeq 5.5~{\rm eV}~,\\[0.2cm]
m_\beta \simeq \sqrt{ \lsndt + |U_{e5}|^2 \, (\lsndt + \lsndth) 
+ |U_{e6}|^2 \, (\lsndt + \lsndo) } \simeq 0.93~{\rm eV}~,\\[0.2cm]
\meff \simeq \sqrt{ \lsndt } \, 
\sqrt{1 - \sin^2 2 \theta_\odot \, \sin^2 \alpha_2/2} 
\simeq (0.36 \div 0.91)~{\rm eV}~.
\eea
Finally, the two heavy sterile 
neutrinos can correspond to \lsndt~and \lsndth~ 
(schemes SSNSc and SSISc): 
\bea 
\Sigma \simeq 3 \sqrt{\lsndo} + 
\sqrt{\lsndo + \lsndth} + \sqrt{\lsndo + \lsndt} 
\simeq 7.2~{\rm eV}~,\\[0.2cm]
m_\beta \simeq \sqrt{ \lsndo + |U_{e5}|^2 \, (\lsndo + \lsndth) 
+ |U_{e6}|^2 \, (\lsndo + \lsndt) } \simeq 1.37~{\rm eV}~,\\[0.2cm]
\meff \simeq \sqrt{ \lsndo } \, 
\sqrt{1 - \sin^2 2 \theta_\odot \, \sin^2 \alpha_2/2} 
\simeq (0.54 \div 1.36)~{\rm eV}~.
\eea
\item[(iv)]the fourth class of mass orderings are scenarios in which 
the two sterile neutrinos are lighter than the active ones 
which in turn are lighter than the last sterile one. As for the 
previous class of scenarios, three possibilities are present. 
In scenarios SNSSa and SISSa the two light sterile neutrinos 
correspond to \lsndt~and \lsndth:
\bea 
\Sigma \simeq 3 \sqrt{\lsndt} + \sqrt{\lsndt - \lsndth} + 
\sqrt{\lsndo + \lsndt }\simeq 5.0~{\rm eV}~,\\[0.2cm]
m_\beta \simeq \sqrt{\lsndt + |U_{e5}|^2 \, (\lsndt - \lsndth) + 
|U_{e6}|^2 \, (\lsndo + \lsndt)} 
\simeq 0.93~{\rm eV}~,\\[0.2cm]
\meff \simeq \sqrt{ \lsndt} \, 
\sqrt{1 - \sin^2 2 \theta_\odot \, \sin^2 \alpha_2/2} 
\simeq (0.36 \div 0.91)~{\rm eV}~.
\eea
If the neutrinos associated with \lsndo~and \lsndth~are 
lighter (scenarios SNSSb and SISSb), then
\bea 
\Sigma \simeq 3 \sqrt{\lsndo} + \sqrt{\lsndo - \lsndth} + 
\sqrt{\lsndo + \lsndt }\simeq 6.9~{\rm eV}~,\\[0.2cm]
m_\beta \simeq \sqrt{\lsndo + |U_{e5}|^2 \, (\lsndo - \lsndth) + 
|U_{e6}|^2 \, (\lsndo + \lsndt)} 
\simeq 1.37~{\rm eV}~,\\[0.2cm]
\meff \simeq \sqrt{ \lsndo} \, 
\sqrt{1 - \sin^2 2 \theta_\odot \, \sin^2 \alpha_2/2} 
\simeq (0.54 \div 1.36)~{\rm eV}~.
\eea
Finally, in scenarios SNSSc and SISSc the heaviest neutrino 
corresponds to \lsndth: 
\bea 
\Sigma \simeq 3 \sqrt{\lsndo} + \sqrt{\lsndo - \lsndt} + 
\sqrt{\lsndo + \lsndth }\simeq 6.6~{\rm eV}~,\\[0.2cm]
m_\beta \simeq \sqrt{\lsndo + |U_{e5}|^2 \, (\lsndo - \lsndt) + 
|U_{e6}|^2 \, (\lsndo + \lsndth)} 
\simeq 1.37~{\rm eV}~,\\[0.2cm]
\meff \simeq \sqrt{ \lsndo} \, 
\sqrt{1 - \sin^2 2 \theta_\odot \, \sin^2 \alpha_2/2} 
\simeq (0.54 \div 1.36)~{\rm eV}~.
\eea
\end{itemize}
Hence, except for scenarios SSSN and SSSI one can expect a signal in 
KATRIN and in neutrinoless double beta decay searches. 
The latter has in this case only three main predictions, given by 
a quasi-degenerate scenario with a common mass scale $\sqrt{\lsndo}$, 
$\sqrt{\lsndt}$ or $\sqrt{\lsndth}$. In order to distinguish these 
cases, conditions in analogy to Eq.~(\ref{eq:cond}) should be fulfilled. 
Note that strictly speaking the analysis of Ref.~\cite{3+2data} 
does not apply to the scenarios discussed in items (iii), (iv) and (v). 
The reason is the same as the one discussed in Section \ref{sec:schemes} 
for schemes SNSa/SISa/SNSb/SISb.

\section{\label{sec:UHE}An alternative non-oscillation probe of Neutrino 
Spectra: Decay of astrophysical Neutrinos}
\renewcommand{\theequation}{B\arabic{equation}}
\setcounter{equation}{0}

Another interesting non-oscillation probe to distinguish different mass 
orderings is the decay of astrophysical neutrinos \cite{decay}. 
Albeit such an analysis depends crucially on the non-trivial assumption 
that neutrinos decay, it is an interesting exercise to investigate 
the implications of such a situation\footnote{For other uses of 
sterile neutrinos in neutrino telescopes, see \cite{st_others}.}. 
Different astrophysical sources can generate a neutrino flux with 
a certain initial composition in $\nu_e$, $\nu_\mu$ and $\nu_\tau$. 
Assuming that all neutrino states 
except for the lightest $\nu_i$ decay, leads to 
\be
\Phi_e : \Phi_\mu : \Phi_\tau = |U_{ei}|^2 : |U_{\mu i}|^2 
: |U_{\tau i}|^2~,
\ee
where $\Phi_\alpha$ with $\alpha = e, \mu, \tau$ is the flux 
of neutrinos 
and anti-neutrinos of flavor $\alpha$ which reaches Earth. 
Note that in the decay scenario this final flux is 
independent on the initial 
flavor composition. The crucial observation is that 
the surviving lightest neutrino state being $\nu_i$, 
the flux $\Phi_\alpha$, which is proportional to $|U_{\alpha i}|^2$,  
may differ in the eight 
neutrino mass orderings under study. 
For scenario SSN we have $i = 1$, while for SSI it holds $i = 3$. 
This corresponds to 
the situation in the three-flavor scenarios studied in 
Refs.~\cite{decay}. In contrast, for scenarios NSS, ISS, SNSb and SISb 
we have $i = 5$, while in the orderings SNSa and SISa it 
holds that $i = 4$. There are therefore four different possibilities, 
and the mass orderings sharing the same phenomenology are in fact the 
same as the ones sharing the same phenomenology of 
the mass-related observables. 

What is eventually measured in neutrino telescopes like 
IceCube \cite{ice} are ratios of fluxes, and here for illustrative 
purposes we will focus on the ratio 
\be 
R_{e\mu} \equiv \frac{\Phi_e}{\Phi_\mu}~,
\ee 
which can be obtained by 
comparing the rate of shower and muon events \cite{emu}. 
Taking ratios including $\nu_\tau$ into account will complicate 
the situation considerably, 
as the mixing elements $|U_{\tau 4}|$ and $|U_{\tau 5}|$ enter the game, 
which are basically unconstrained (one could however use 
these ratios to obtain information on these elements). 
We assume maximal atmospheric neutrino mixing, 
take the best-fit values from Eqs.~(\ref{eq:data}, \ref{eq:stBF}) 
and use from Ref.~\cite{3+2data} that $|U_{\mu 5}| = 0.12$ 
as well as $|U_{\mu 4}| = 0.16$. 
It follows for the ratios that 
$R_{e\mu}^{\rm SSN} = 2/\tan^2 \theta_\odot = 4.7$, 
$R_{e\mu}^{\rm SSI} = 2 \sin^2 \theta_{\rm CHOOZ} = 0$, 
$R_{e\mu}^{\rm NSS, \, ISS, \, SNSb, \, SISb} 
= |U_{e 5}|^2/|U_{\mu 5}|^2 = 1$ 
and 
$R_{e\mu}^{\rm SNSa, \, SISa} = |U_{e 4}|^2/|U_{\mu 4}|^2 = 0.47$. 
The ratios are easily distinguishable from each other, in particular 
SSN and SSI, which we have shown in Section \ref{sec:3l2h} to be 
very similar in the mass-related observables. Note however that 
for standard astrophysical sources (an initial composition of 
$1 : 2 : 0$) and no decay the ratio $R_{e\mu}$ is equal to 1 
for maximal atmospheric mixing and $U_{e3}=0$, i.e., identical 
to $R_{e\mu}^{\rm NSS, \, ISS, \, SNSb, \, SISb}$. 
In addition, taking the uncertainty of the mixing 
matrix elements into account 
complicates the situation further. Taking the $3\sigma$ ranges 
from Ref.~\cite{GonMal} (in particular 
$\sin^2 \theta_{\rm A} = (0.32 \div 0.64)$ for atmospheric 
neutrino mixing) and assuming again a 50\% uncertainty on 
the sterile neutrino parameters gives that 
$R_{e\mu}^{\rm SSN} = (2.2 \div 10.8)$, 
$R_{e\mu}^{\rm SSI} = (0 \div 0.13)$, 
$R_{e\mu}^{\rm NSS, \, ISS, \, SNSb, \, SISb} = (0.11 \div 9.0)$ 
and $R_{e\mu}^{\rm SNSa, \, SISa} = (0.06 \div 4.0)$. 
The standard scenario predicts $R_{e \mu}$ between 0.73 and 1.19, 
a range which is covered by all decay scenarios except for SSN and SSI. 
The ratios are now overlapping and except for the cases of measuring 
very small ($R_{e \mu} \le 0.06$) or large ratios 
($R_{e \mu} \ge 9$) no possibility can be unambiguously 
identified. However, certain cases can be ruled out, for instance 
scenarios SSI, SNSa and SISa for an observation of 
$R_{e \mu} \ge 4$. 
To fully disentangle the different cases the mixing 
parameters the mixing matrix elements should be known 
much more precisely. 

\end{appendix}


\SetWidth{1.23}
\begin{center}
\begin{figure}\hspace{5cm}
\begin{picture}(300,400)(0,0)
\Line(-80,360)(45,360)
\Line(-80,280)(45,280)
\LongArrow(-05,58)(-05,23)
\LongArrow(-05,23)(-05,58)
\Text(-45,42)[l]{\large \dma}
\Line(-80,60)(45,60)
\LongArrow(-05,02)(-05,17)
\LongArrow(-05,17)(-05,02)
\Text(-45,10)[l]{\large \dms}
\Line(-80,20)(45,20)
\Line(-80,0)(45,0)
\LongArrow(50,358)(50,2)
\LongArrow(50,2)(50,358)
\LongArrow(25,278)(25,2)
\LongArrow(25,2)(25,278)
\Text(-90,362)[l]{\large 5}
\Text(-90,282)[l]{\large 4}
\Text(-90,62)[l]{\large 3}
\Text(-90,22)[l]{\large 2}
\Text(-90,2)[l]{\large 1}
\rText(26,170)[l][r]{\large \lsndo}
\rText(0,170)[l][r]{\large \lsndt}
\Text(-20,-30)[c]{\Huge \bf SSN}
\Line(130,360)(255,360)
\Line(130,280)(255,280)
\LongArrow(205,38)(205,03)
\LongArrow(205,03)(205,38)
\Text(165,22)[l]{\large \dma}
\Line(130,60)(255,60)
\LongArrow(205,42)(205,57)
\LongArrow(205,57)(205,42)
\Text(165,50)[l]{\large \dms}
\Line(130,40)(255,40)
\Line(130,0)(255,0)
\LongArrow(260,358)(260,2)
\LongArrow(260,2)(260,358)
\LongArrow(235,278)(235,2)
\LongArrow(235,2)(235,278)
\Text(120,362)[l]{\large 5}
\Text(120,282)[l]{\large 4}
\Text(120,62)[l]{\large 2}
\Text(120,42)[l]{\large 1}
\Text(120,2)[l]{\large 3}
\rText(236,170)[l][r]{\large \lsndo}
\rText(210,170)[l][r]{\large  \lsndt}
\Text(190,-30)[c]{\Huge \bf SSI}
\end{picture}
\vspace{1.8cm}
\caption{\label{fig:3l2h}Allowed 2+3 mass orderings which are 
defined by having two sterile neutrinos heavier than the three active 
neutrinos. (Not to scale).}
\end{figure}
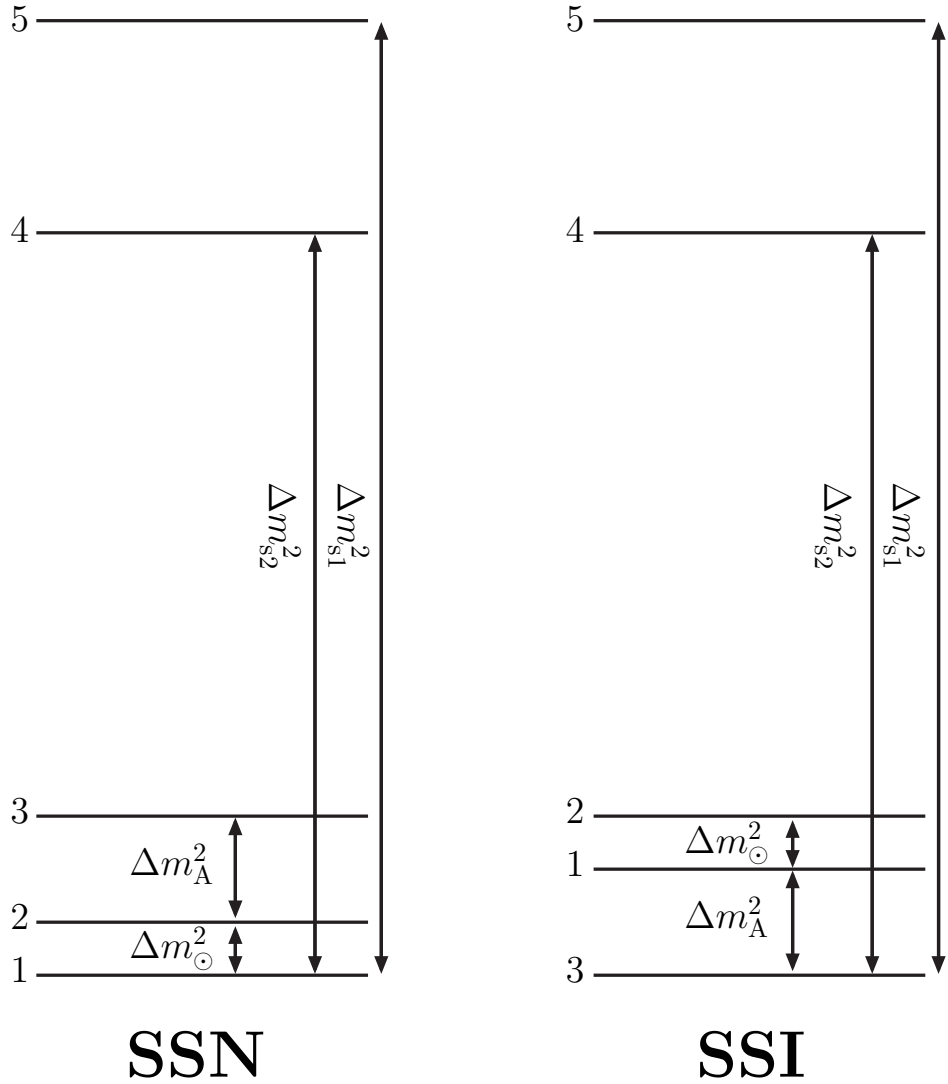
\end{center}

\SetWidth{1.23}
\begin{center}
\begin{figure}\hspace{5cm}
\begin{picture}(300,400)(0,0)
\Line(-80,360)(45,360)
\Line(-80,320)(45,320)
\LongArrow(-5,358)(-5,323)
\LongArrow(-5,323)(-5,358)
\Text(-45,342)[l]{\large \dma}
\Line(-80,300)(45,300)
\LongArrow(-5,302)(-5,317)
\LongArrow(-5,317)(-5,302)
\Text(-45,311)[l]{\large \dms}
\Line(-80,80)(45,80)
\Line(-80,0)(45,0)
\LongArrow(50,358)(50,2)
\LongArrow(50,2)(50,358)
\LongArrow(25,318)(25,2)
\LongArrow(25,2)(25,318)
\Text(-90,362)[l]{\large 3}
\Text(-90,322)[l]{\large 2}
\Text(-90,302)[l]{\large 1}
\Text(-90,82)[l]{\large 4}
\Text(-90,2)[l]{\large 5}
\rText(26,180)[l][r]{\large \lsndo}
\rText(00,180)[l][r]{\large  \lsndt}
\Text(-20,-30)[c]{\Huge \bf NSS}
\Line(130,360)(255,360)
\Line(130,340)(255,340)
\LongArrow(205,358)(205,343)
\LongArrow(205,343)(205,358)
\Text(165,351)[l]{ \large \dms}
\Line(130,300)(255,300)
\LongArrow(205,302)(205,337)
\LongArrow(205,337)(205,302)
\Line(130,80)(255,80)
\Text(165,322)[l]{\large \dma}
\Line(130,0)(255,0)
\LongArrow(260,358)(260,2)
\LongArrow(260,2)(260,358)
\LongArrow(235,82)(235,358)
\LongArrow(235,358)(235,82)
\Text(120,362)[l]{\large 2}
\Text(120,342)[l]{\large 1}
\Text(120,302)[l]{\large 3}
\Text(120,82)[l]{\large 4}
\Text(120,2)[l]{\large 5}
\rText(236,180)[l][r]{\large \lsndo}
\rText(210,180)[l][r]{\large \lsndt}
\Text(190,-30)[c]{\Huge \bf ISS}
\end{picture}
\vspace{1.8cm}
\caption{\label{fig:3h2l}
Allowed 3+2 mass orderings which are 
defined by having two sterile neutrinos lighter than the three active 
neutrinos. (Not to scale).} 
\end{figure}
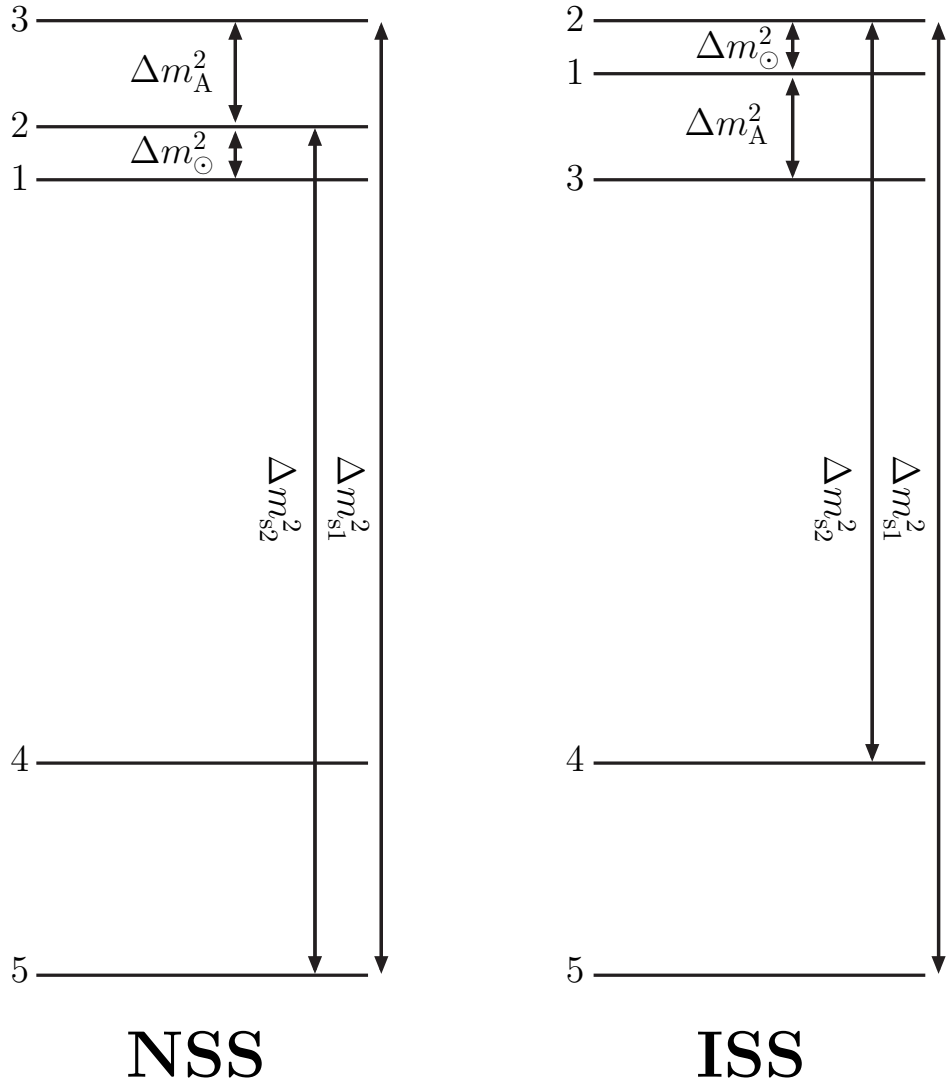
\end{center}

\begin{samepage}
\SetWidth{1.23}
\begin{center}
\begin{figure}\hspace{0.3cm}
\begin{picture}(600,400)(0,0)
\Line(-60,360)(45,360)
\Line(-60,0)(45,0)
\LongArrow(-5,208)(-5,173)
\LongArrow(-5,173)(-5,208)
\Text(-45,192)[l]{\large \dma}
\Line(-60,210)(45,210)
\LongArrow(-5,152)(-5,167)
\LongArrow(-5,167)(-5,152)
\Text(-45,160)[l]{\large \dms}
\Line(-60,170)(45,170)
\Line(-60,150)(45,150)
\LongArrow(50,358)(50,152)
\LongArrow(50,152)(50,358)
\LongArrow(50,148)(50,2)
\LongArrow(50,2)(50,148)
\Text(-70,362)[l]{\large 5}
\Text(-70,2)[l]{\large 4}
\Text(-70,212)[l]{\large 3}
\Text(-70,172)[l]{\large 2}
\Text(-70,152)[l]{\large 1}
\rText(26,290)[l][r]{\large \lsndot}
\rText(26,80)[l][r]{\large \lsndtt}
\Text(-7,-30)[c]{\Huge \bf SNSa}
\Line(85,360)(190,360)
\Line(85,0)(190,0)
\LongArrow(140,268)(140,233)
\LongArrow(140,233)(140,268)
\Text(100,252)[l]{\large \dma}
\Line(85,210)(190,210)
\LongArrow(140,212)(140,227)
\LongArrow(140,227)(140,212)
\Text(100,220)[l]{\large \dms}
\Line(85,230)(190,230)
\Line(85,270)(190,270)
\LongArrow(195,358)(195,212)
\LongArrow(195,212)(195,358)
\LongArrow(195,208)(195,2)
\LongArrow(195,2)(195,208)
\Text(75,362)[l]{\large 4}
\Text(75,2)[l]{\large 5}
\Text(75,272)[l]{\large 3}
\Text(75,232)[l]{\large 2}
\Text(75,212)[l]{\large 1}
\rText(171,310)[l][r]{\large \lsndtt}
\rText(171,100)[l][r]{\large  \lsndot}
\Text(136,-30)[c]{\Huge \bf SNSb}
\Line(230,360)(335,360)
\Line(230,0)(335,0)
\LongArrow(285,188)(285,153)
\LongArrow(285,153)(285,188)
\Text(245,172)[l]{\large \dma}
\Line(230,210)(335,210)
\LongArrow(285,192)(285,207)
\LongArrow(285,207)(285,192)
\Text(245,200)[l]{\large \dms}
\Line(230,190)(335,190)
\Line(230,150)(335,150)
\LongArrow(340,358)(340,152)
\LongArrow(340,152)(340,358)
\LongArrow(340,148)(340,2)
\LongArrow(340,2)(340,148)
\Text(220,362)[l]{\large 5}
\Text(220,2)[l]{\large 4}
\Text(220,212)[l]{\large 2}
\Text(220,192)[l]{\large 1}
\Text(220,152)[l]{\large 3}
\rText(316,290)[l][r]{\large \lsndot}
\rText(316,80)[l][r]{\large  \lsndtt}
\Text(279,-30)[c]{\Huge \bf SISa}
\Line(375,360)(480,360)
\Line(375,0)(480,0)
\LongArrow(430,248)(430,213)
\LongArrow(430,213)(430,248)
\Text(390,232)[l]{\large \dma}
\Line(375,270)(480,270)
\LongArrow(430,252)(430,267)
\LongArrow(430,267)(430,252)
\Text(390,260)[l]{\large \dms}
\Line(375,250)(480,250)
\Line(375,210)(480,210)
\LongArrow(485,358)(485,212)
\LongArrow(485,212)(485,358)
\LongArrow(485,206)(485,2)
\LongArrow(485,2)(485,205)
\Text(365,362)[l]{\large 4}
\Text(365,2)[l]{\large 5}
\Text(365,272)[l]{\large 2}
\Text(365,252)[l]{\large 1}
\Text(365,212)[l]{\large 3}
\rText(461,310)[l][r]{\large \lsndtt}
\rText(461,100)[l][r]{\large  \lsndot}
\Text(424,-30)[c]{\Huge \bf SISb}
\end{picture}
\vspace{1cm}
\caption{\label{fig:131}Allowed 1+3+1 mass orderings which 
are defined by having one sterile neutrino heavier than the 
three active ones 
which in turn are heavier than the second 
sterile neutrino. (Not to scale). Note that not necessarily 
$\lsndot = \lsndo$ and $\lsndtt = \lsndt$ holds.}
\end{figure}
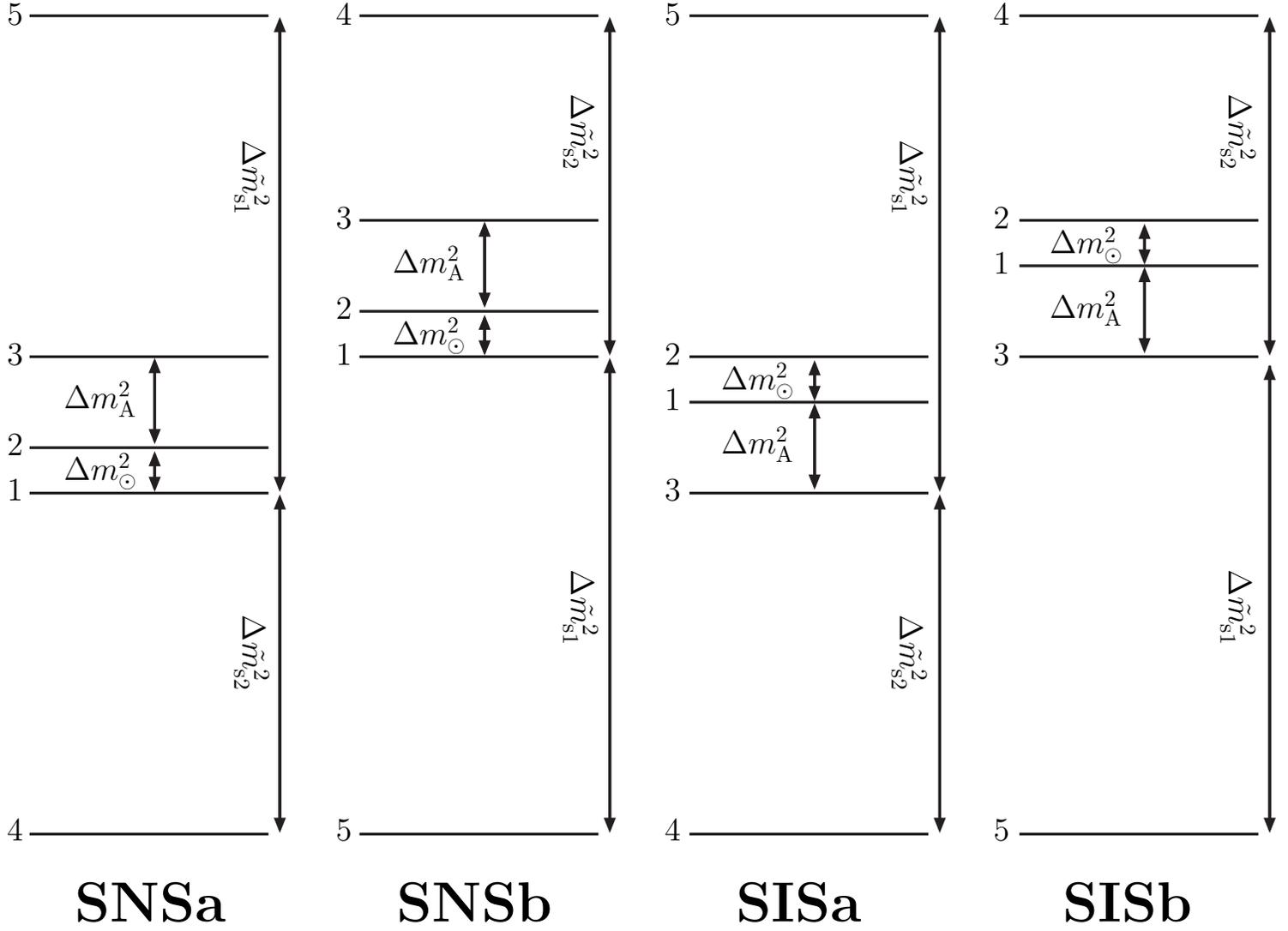
\end{center}

\end{samepage}


\begin{figure} \vspace{-.3cm}
\epsfig{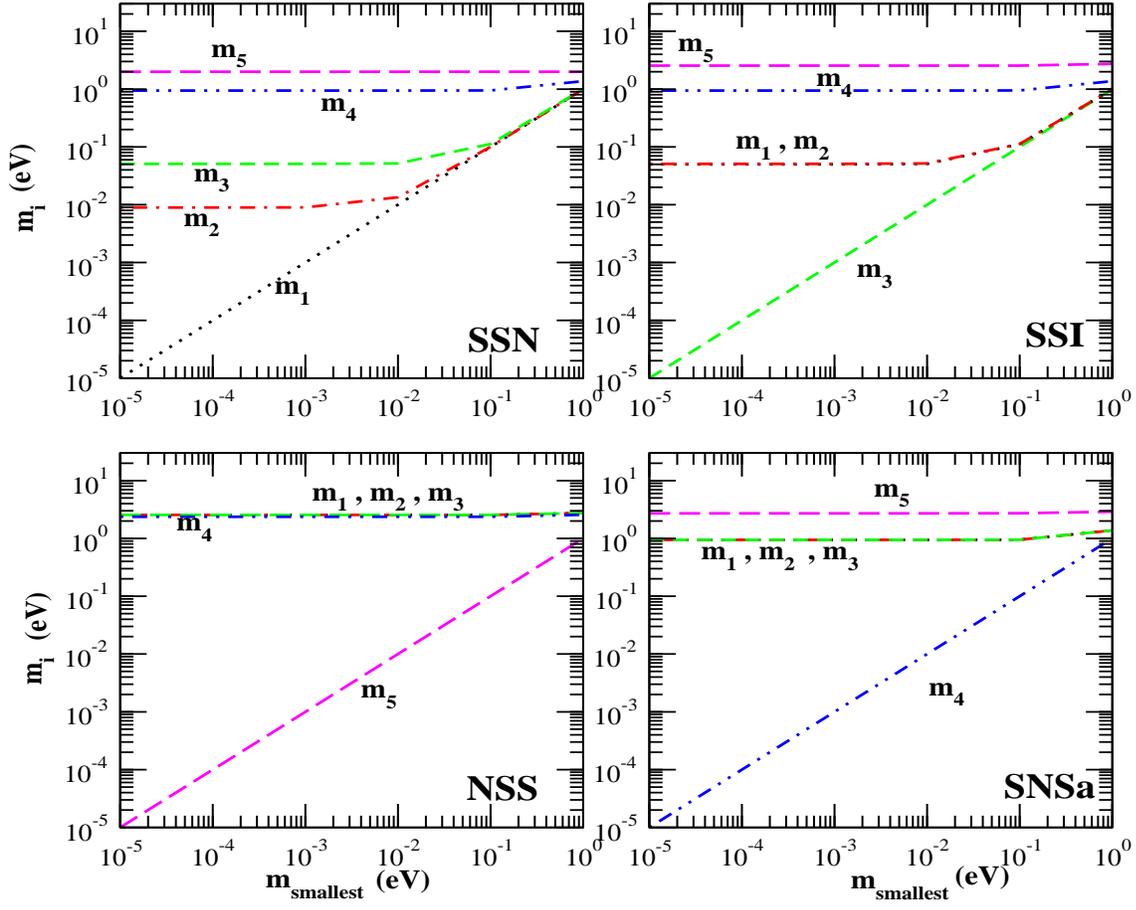}
\caption{\label{fig:masses}
Scenarios with three active and two sterile neutrinos: 
the individual neutrino masses as a function of the smallest neutrino 
mass for scenarios SSN, SSI, NSS and SNSa. 
For the mass-squared differences related to the 
sterile neutrinos the best-fit point given in Eq.~(\ref{eq:stBF}) 
is used and we assumed that $\lsndo = \lsndot$ and $\lsndt = \lsndtt$. 
Scenario ISS is indistinguishable from case NSS and SISa from 
SNSa. The schemes SNSb and SISb are very similar to NSS.}
\end{figure}


\begin{figure}
\epsfig{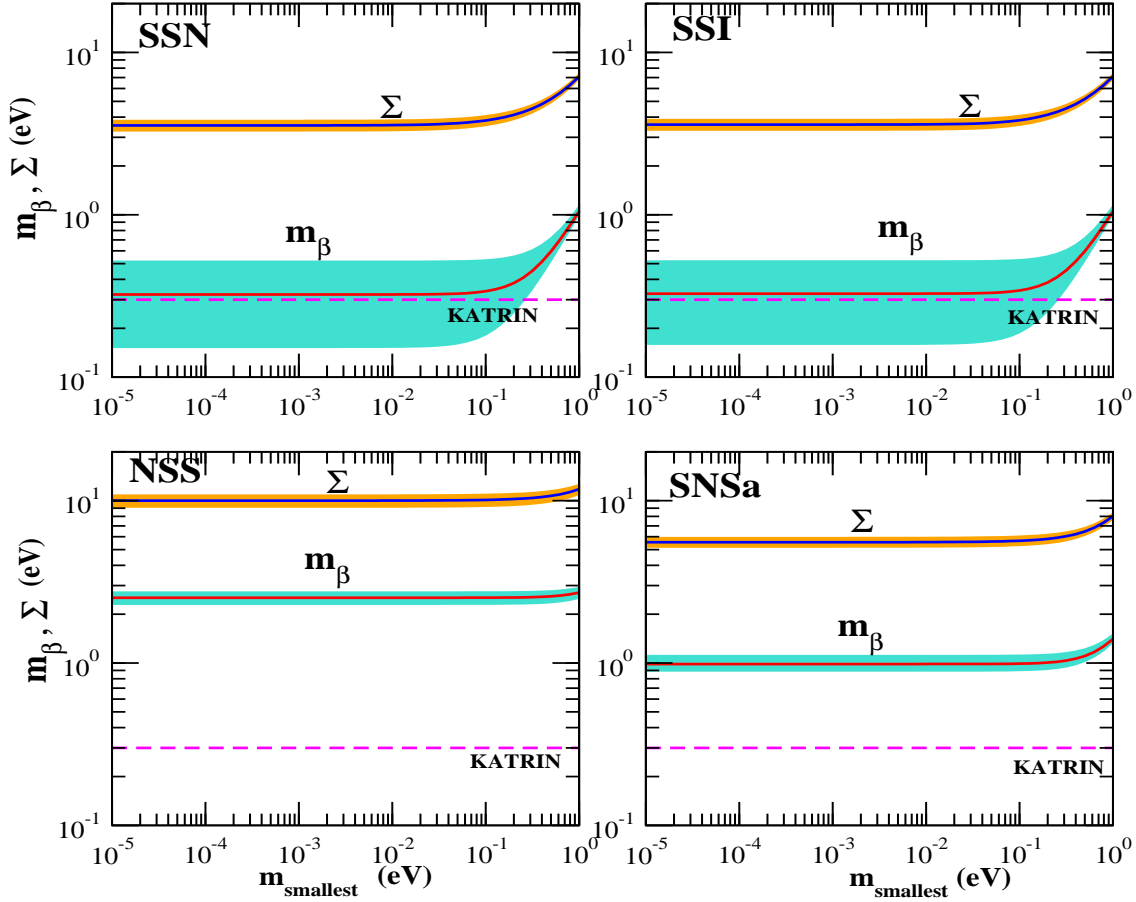}  
\caption{\label{fig:mass1}The sum of neutrino masses $\Sigma$ and the 
kinematic neutrino mass $m_\beta$ for scenarios SSN (top left), SSI 
(top right), NSS (bottom left) and SNSa (bottom right). 
The solid lines give the values of the respective observable 
at the 
best-fit point Eq.~(\ref{eq:stBF})while the shaded regions are obtained by 
varying the parameters involved 
in  their corresponding 
ranges from Eqs.~(\ref{eq:data}) and (\ref{eq:rangeUe45}). 
We assumed that $\lsndo = \lsndot$ and $\lsndt = \lsndtt$. 
Scenario ISS is indistinguishable from case NSS, 
SISa is indistinguishable from SNSa, and SNSb/SISb are 
indistinguishable from NSS. 
For these two observables SSN and SSI give identical results. 
Also indicated is the KATRIN sensitivity on $m_\beta$ of 0.3 eV.}
\end{figure}


\begin{figure}
\epsfig{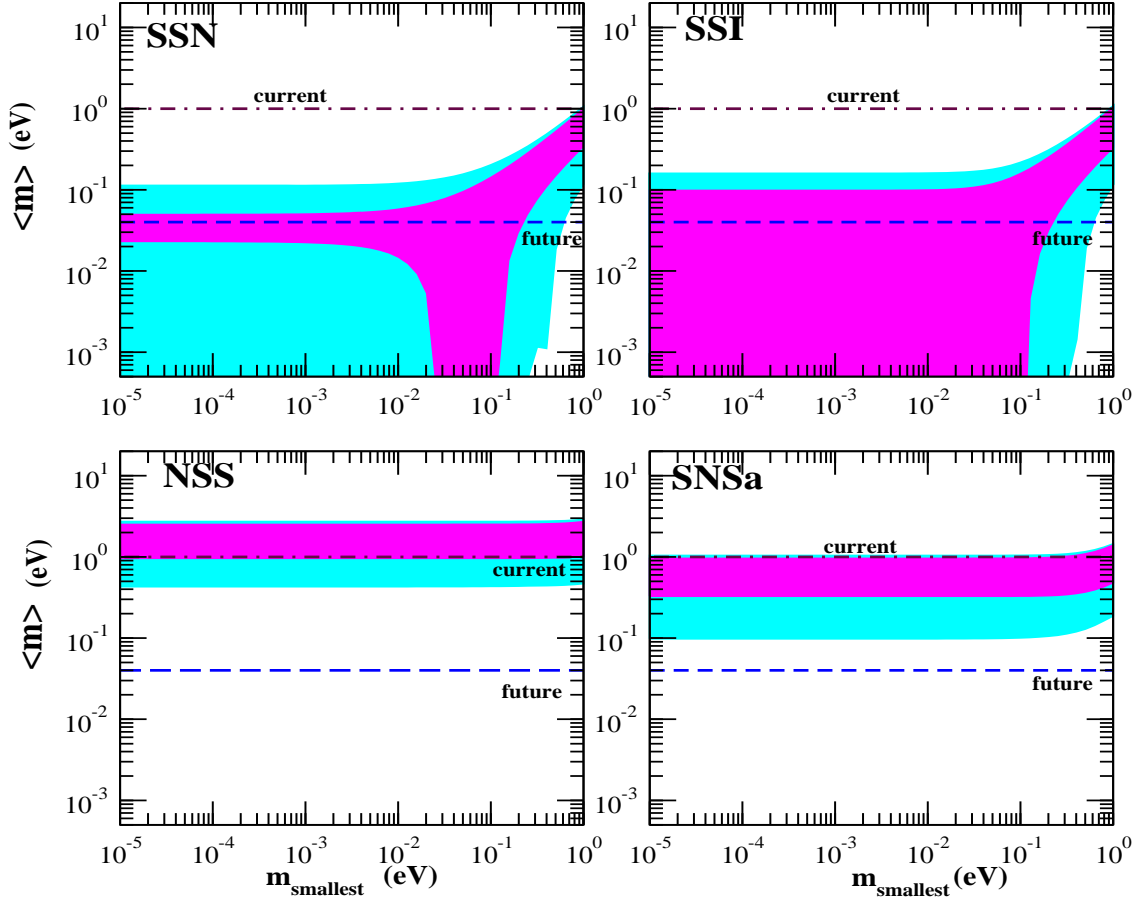}  
\caption{\label{fig:meff1}The effective mass \meff~for scenarios 
SSN (top left), SSI 
(top right), NSS (bottom left) and SNSa (bottom right). 
The magenta shaded (darker) regions correspond to the 
the best-fit point from Eq.~(\ref{eq:stBF}) and allowing the 
phases to take arbitrary values in the interval [0 : 2$\pi$]. 
The blue shaded (lighter) regions are obtained by varying the 
mass and mixing angles as well in 
the allowed ranges from Eqs.~(\ref{eq:data}) and (\ref{eq:rangeUe45}). 
We assumed that $\lsndo = \lsndot$ and $\lsndt = \lsndtt$. 
Scenario ISS is indistinguishable from case NSS, 
SISa is indistinguishable from SNSa, and SNSb/SISb are 
indistinguishable from NSS. Also indicated is the current limit of 1 eV 
and a future bound of 0.04 eV.}
\end{figure}

\begin{figure}
\epsfig{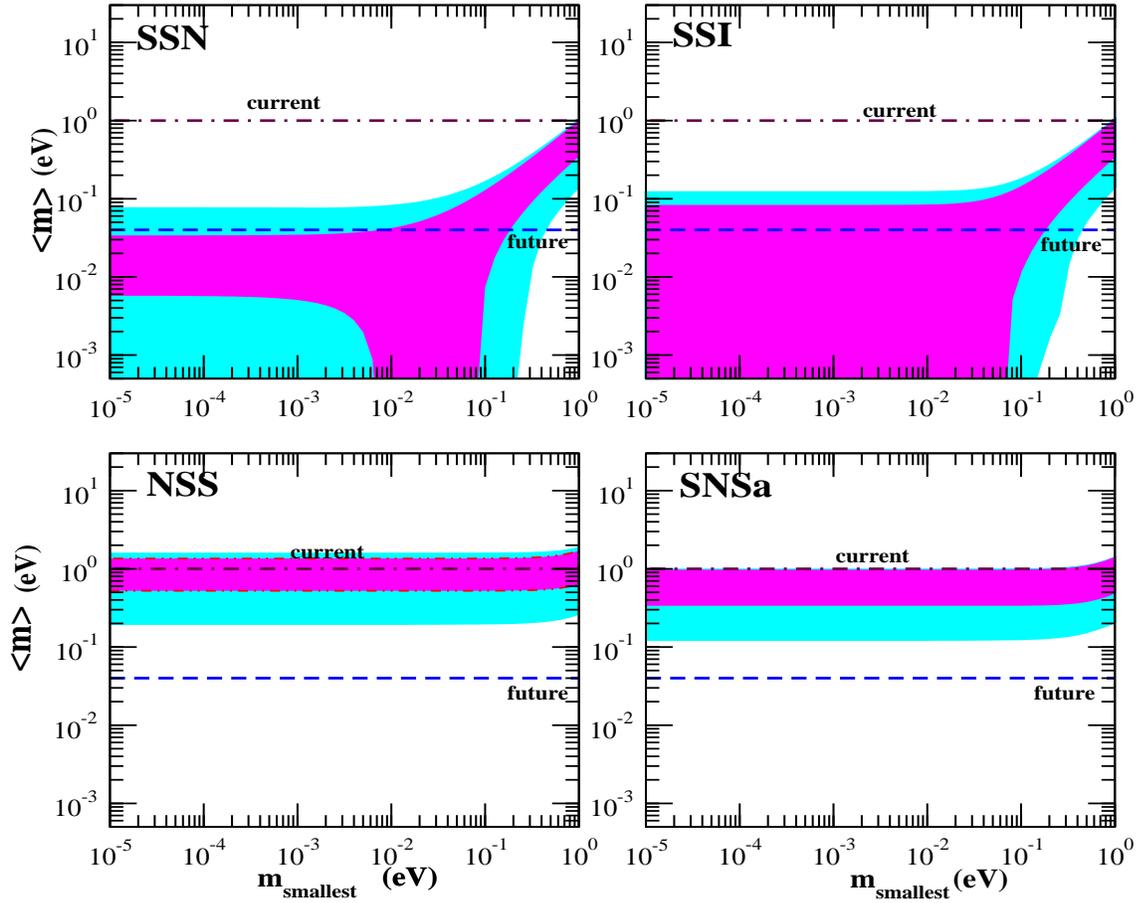}
\caption{\label{fig:meff2}Same as Fig.~\ref{fig:meff1} for the second 
typical sterile parameter point 
and the corresponding range from Eq.~(\ref{eq:st2}).}
\end{figure}

\begin{figure}
\epsfig{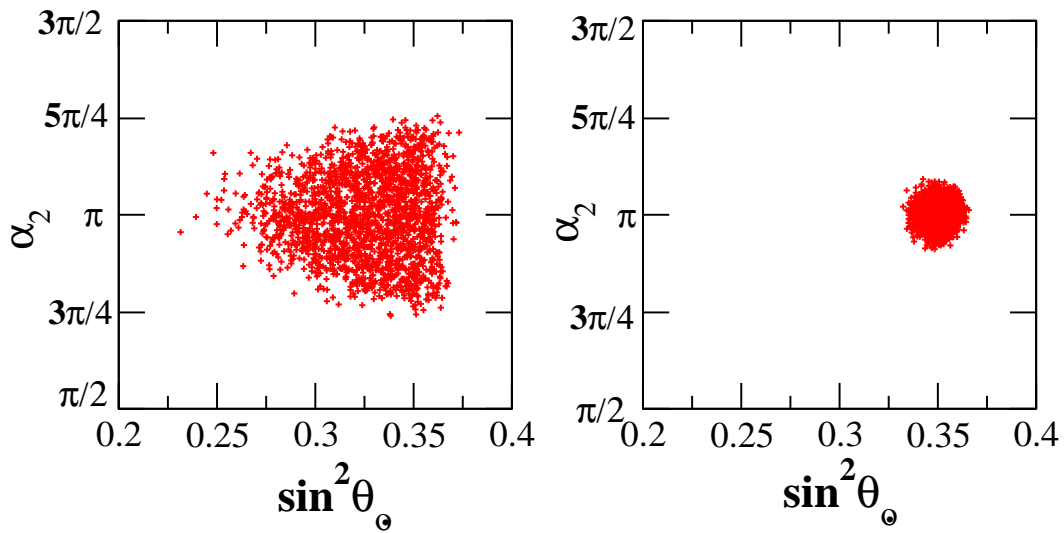}  
\caption{\label{fig:meff_exa}Allowed values of the Majorana 
phase $\alpha_2$ and 
$\sin^2 \theta_\odot$ in scenarios NSS and ISS for the current 
limit on \meff~of 
1 eV (left) and a future limit of 0.5 eV (right). 
The 3$\sigma$ ranges from 
Eq.~(\ref{eq:data}) and the ranges around the best-fit point 
from Eq.~(\ref{eq:stBF}) are used.}
\end{figure}

\end{document}